\documentclass[journal,final,twoside]{IEEEtran}
\usepackage{cite}

\ifCLASSINFOpdf
   \usepackage[pdftex]{graphicx}
   \graphicspath{{./figures/}{./pdf/}{./jpeg/}}
   \DeclareGraphicsExtensions{.pdf,.jpeg,.png}
\else
   \usepackage[dvips]{graphicx}
   \graphicspath{{./eps/}}
   \DeclareGraphicsExtensions{.eps}
\fi
\usepackage{algorithm}
\usepackage{algpseudocode}

\usepackage[cmex10]{amsmath}
\usepackage{amssymb}
\usepackage{hyperref}

\usepackage[caption=false,font=footnotesize]{subfig}
\usepackage{stfloats}

\usepackage{multirow}

\usepackage{footnote}

\usepackage{threeparttable}

\usepackage{xcolor}

\renewcommand{\vec}[1]{\mathbf{#1}}

\newcommand{\vecgreek}[1]{\boldsymbol{#1}}

\newcommand{\mtx}[1]{\mathbf{#1}}

\begin{document}
%
\title{Regularized Compression of MRI Data: Modular Optimization of Joint Reconstruction and Coding}
%
%
%

\author{Veronica Corona, Yehuda Dar, Guy Williams, and Carola-Bibiane Sch{\"o}nlieb 
\\
\thanks{V.~Corona and C.-B.~Sch{\"o}nlieb are with the Department of Applied Mathematics and Theoretical Physics, University of Cambridge, U.K.  E-mail addresses: \{vc324,~cbs31\}@cam.ac.uk.}
\thanks{Y.~Dar is with the Department of Electrical and Computer Engineering, Rice University, U.S.  E-mail address: ydar@rice.edu.}
\thanks{G.~Williams is with the Department of Clinical Neurosciences, University of Cambridge, U.K.  E-mail address: gbw1000@cam.ac.uk.}
}

%
%

\markboth{}%
{~}
%



\maketitle

\begin{abstract}
The Magnetic Resonance Imaging (MRI) processing chain starts with a critical acquisition stage that provides raw data for reconstruction of images for medical diagnosis. This flow usually includes a near-lossless data compression stage that enables digital storage and/or transmission in binary formats. In this work we propose a framework for joint optimization of the MRI reconstruction and lossy compression, producing compressed representations of medical images that achieve improved trade-offs between quality and bit-rate. 
Moreover, we demonstrate that lossy compression can even improve the reconstruction quality compared to settings based on lossless compression. 
Our method has a modular optimization structure, implemented using the alternating direction method of multipliers (ADMM) technique and the state-of-the-art image compression technique (BPG) as a black-box module iteratively applied. This establishes a medical data compression approach compatible with a lossy compression standard of choice.  A main novelty of the proposed algorithm is in the total-variation regularization added to the modular compression process, leading to decompressed images of higher quality without any additional processing at/after the decompression stage. 
Our experiments show that our regularization-based approach for joint MRI reconstruction and compression often achieves significant PSNR gains between 4 to 9 dB at high bit-rates compared to non-regularized solutions of the joint task. 
Compared to regularization-based solutions, our optimization method provides PSNR gains between 0.5 to 1 dB at high bit-rates, which is the range of interest for medical image compression. 

\end{abstract}


~\\~

%
\IEEEpeerreviewmaketitle

\section{Introduction}
\begin{figure*}[]
	\centering
	{\subfloat[(a) Decoupled reconstruction and compression]{\label{fig:decoupled_process}\includegraphics[width=0.9\textwidth]{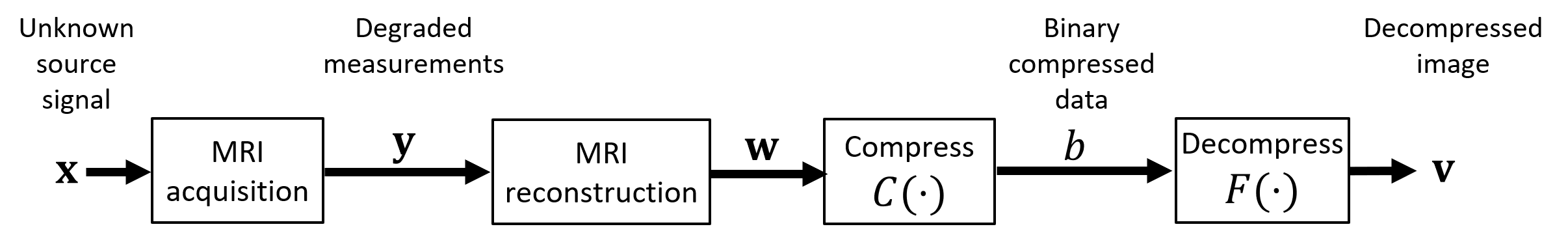}}}
	\\
	{\subfloat[(b) Joint reconstruction and compression]{\label{fig:joint_process}\includegraphics[width=0.9\textwidth]{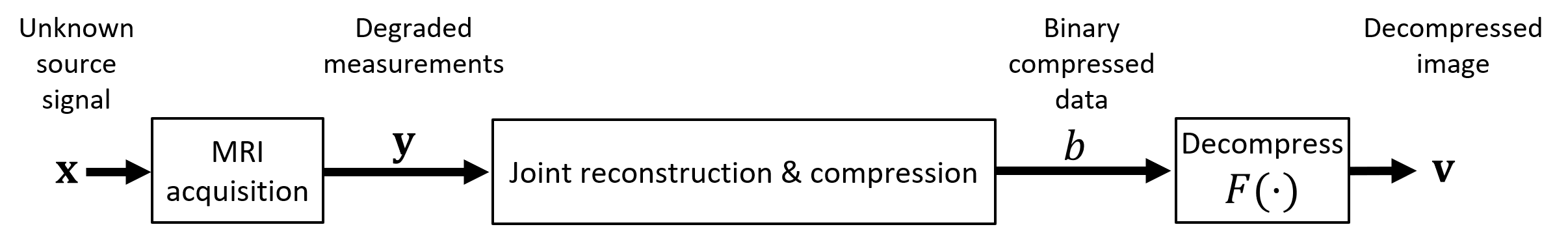}}}
	\caption{Two general settings for an MRI processing system: (a) based on a decoupled reconstruction and compression stages, and (b) based on a joint reconstruction and compression stage.} 
\end{figure*}

\IEEEPARstart{A}{cquisition} and compression are two crucial stages in imaging systems. The acquired data, describing the visual scene in a degraded form, is further processed to reconstruct images that can be compressed for storage or transmission. The decompressed images are the important outcome of this overall process, and ideally should be of a good quality considering the bit-rate required for their binary compressed representations. In practice, this ultimate goal of optimizing the entire processing chain is often neglected, leaving the individual tasks unaware of each other (Fig.~\autoref{fig:decoupled_process}) and, by that, inducing sub-optimal performance with respect to the complete system. 


Recently there was a significant progress in joint optimization of multiple tasks within typical imaging systems. This is mainly due to the emerging utilization of optimization techniques relying on variable splitting methods (e.g., the alternating direction method of multipliers (ADMM) \cite{boyd2011distributed}), promoting numerically tractable designs that improved the state-of-the-art performance for various multi-task goals. 

Contemporary examples for the benefits of multi-task optimization in the context of MRI include joint reconstruction and segmentation \cite{corona2019enhancing,adler2018task}, reconstruction and registration \cite{corona2019multi,Royuela-del-Val2016,Lingala::2015}, as well as reconstruction and motion estimation \cite{odille2016joint,Rank20174DRM,aviles2018compressed}. The idea in these joint models is to exploit complementary information coming from the different imaging tasks to 
enhance the image reconstruction quality and finally improve the overall performance of the joint optimization. By careful modeling and coupling of the multiple tasks, the authors show accuracy boost as well as reduction of error propagation.

Another recent research line addresses multi-task problems involving image and video compression \cite{dar2018optimized,dar2018system,dar2018compression,dar2019benefiting}. This new design philosophy enables optimization of
compression with respect to complete acquisition-rendering systems \cite{dar2018optimized,dar2018system,dar2018compression}, multimedia
distribution networks \cite{dar2018compression}, and reliable storage systems \cite{dar2019benefiting}.
These architectures are based on ADMM and also have the important property of modularity, where a repeated sub-problem in the iterative optimization process is identified as a standard image compression task and, accordingly, replaced with a repeated black-box application of an existing compression method. The improved compression ability (namely, the better rate-distortion trade-offs achieved) is provided in binary compressed data compatible with standard decompression processes that do not require additional post-decompression processing. 

This modular optimization approach for compression  \cite{dar2018optimized,dar2018system,dar2018compression,dar2019benefiting} relates to the Plug-and-Play Priors framework \cite{venkatakrishnan2013plug,sreehari2016plug}, addressing image reconstruction and restoration problems using ADMM-based iterative procedures that employ denoisers as black-box modules. This utilization of denoisers is motivated by associating a sub-problem in the process with the task of denoising an image contaminated by additive white Gaussian noise.  Similar benefits of denoising-based modularity were also proposed by additional architectures employing optimization techniques beyond ADMM, a prominent example is the Regularization-by-Denoising (RED) framework \cite{romano2017little}, also suggesting powerful solutions to intricate restoration tasks based on iterative applications of state-of-the-art denoisers. This recent branch of research attracts great interest, as reflected in a long line of studies presenting new modular algorithms for restoration \cite{venkatakrishnan2013plug,sreehari2016plug,romano2017little,dar2016postprocessing,rond2016poisson,dar2018restoration} and
compression \cite{dar2018optimized,dar2018system,dar2018compression,dar2019benefiting} tasks. 

Lossy compression of magnetic resonance (MR) images is a topic of current interest. For example, see the overviews given in recent papers \cite{liu2019fast,kumar2020versatile}. 
Compatibility with existing image compression standards is also desired. For example, note the wavelet-based extension proposed for JPEG2000 in \cite{bruylants2015wavelet}. Also note the region-of-interest based method in \cite{yee2017medical} that is compatible with the Better Portable Graphics (BPG) state-of-the-art image compression format. 
As explained next, our approach is also compatible with existing image compression standards. Also, in contrast to the existing studies that focus on the single task of MRI lossy compression, the main contribution of this paper is a framework for joint optimization of MRI reconstruction and lossy compression.

In this paper we propose a new modular optimization approach that brings together the tasks of MRI reconstruction and lossy data compression (see Fig.~\autoref{fig:joint_process}). Specifically, given raw MRI measurements obtained from the acquisition stage, we use the ADMM-based compression approach from \cite{dar2018optimized,dar2018system,dar2018compression,dar2019benefiting} in conjunction with the relevant MRI acquisition model and obtain a process that jointly optimizes the tasks of MRI reconstruction and the lossy compression of the reconstructed MR image. The resulting binary compressed data is compatible with an image compression standard of choice that is employed as a black box in the modular optimization process. 

Moreover, this paper is the first to propose a modular optimization process that includes an explicit, common regularizer (total variation \cite{rudin1992nonlinear} in our case) in modular optimization for lossy compression purposes. This essentially improves the generic image model used in the standard compression by adjusting it to image models that better suit the considered image types. Nicely, this modular optimization still provides compressed data compatible to the compression standard and does not require additional post-decompression processing. 

We present experiments that consider a noisy linear model for MRI acquisition at several levels 
of subsampling of coefficients in the Fourier domain (also known as the k-space in the context of MRI). Our experiments show how the state-of-the-art image compression technique, BPG \cite{hevc_software_bpg}, is adjusted and improved for the purpose of MRI reconstruction and compression. 
We examine two settings of the proposed method: joint optimization of MRI reconstruction and lossy compression \textit{without any additional regularization}, and joint optimization of MRI reconstruction and lossy compression \textit{with total-variation regularization}. Note that the added total-variation regularization is only active at the compression optimization stage and not at (nor after) the decompression stage. 
Our experiments show that our joint optimization approach \textit{with total-variation regularization} often achieves significant PSNR gains between 4 to 9 dB at high bit-rates compared to the joint optimization approach \textit{without any additional regularization}. 
Moreover, our joint optimization approach \textit{with total-variation regularization} provides PSNR gains between 0.5 to 1 dB at high bit-rates compared to a decoupled approach (Fig.~\ref{fig:decoupled_process}) where total-variation based MRI reconstruction is followed by standard lossy compression without any joint optimization process.
These impressive PSNR gains at high bit-rates are of great relevance for medical image compression. 
In addition, we show that lossy compression can significantly improve the MRI reconstruction quality even with respect to total-variation based reconstruction without any compression (i.e., an ideal lossless compression).

This paper is organized as follows. In Section \ref{sec:The Proposed Method} we present the development of the modular optimization approach for joint  reconstruction and compression of MRI data. In Section \ref{sec:Experimental Results} we provide the experimental results. Section \ref{sec:Conclusion} concludes this paper.

\section{The Proposed Method}
\label{sec:The Proposed Method}

\subsection{Problem Formulation}
\label{subsec:Problem Formulation}

We consider an imaging system model (see  Fig.~\ref{fig:joint_process}) where an unknown visual signal, denoted as the $ N $-length column-vector $ \vec{x} \in \mathbb{R}^N $, goes through an MRI acquisition stage modeled via 
\begin{IEEEeqnarray}{rCl}
\label{eq:MRI acquisition model}
	\label{eq:network structure - reconstructed k^th output}
	\vec{y}  = \mtx{A} \vec{x} + \vecgreek{\eta} .
\end{IEEEeqnarray}
Here $\mtx{A}$ is a $M\times N$ matrix that acts as a linear operator that subsamples $\vec{x}$ in the discrete Fourier domain (also known as the k-space in the context of MRI) to return ${M \le N}$ measurements of the visual scene embodied in $\vec{x}$. Also, ${\vecgreek{\eta}\in \mathbb{R}^M}$ is a Gaussian noise vector with i.i.d.~components that are zero mean and have variance $\sigma_{\eta}^2$. Then, $\vec{y}$ is a column vector containing $M$ degraded samples in the discrete Fourier domain. The data in $\vec{y}$ should be reconstructed in order to obtain a visually-meaningful image. 

The linear operator in the acquisition can be formulated as  $\mtx{A}\triangleq\mtx{S}\mtx{F}$, where $\mtx{F}$ is the Discrete Fourier Transform matrix of $N\times N$ size, and $\mtx{S}$ is a subsampling operator in the form of a $M\times N$ matrix that selects $M\le N$ samples from its $N$-dimensional input based on a desired binary sampling pattern. The subsampling ratio $N/M$ defines the MRI acceleration factor, e.g., using acceleration factor 4x keeps and utilizes only 25\% of the coefficients in the Fourier domain. 
Also note that since we consider a true image $\vec{x}$ which is real valued, we can adapt $\mtx{A}$ and its adjoint $\mtx{A}^{*}$ to be $\mathbb{R}^{N}\rightarrow\mathbb{C}^{M}$ and $\mathbb{C}^{M}\rightarrow\mathbb{R}^{N}$ linear operators, respectively, as described in \cite{ehrhardt2016multicontrast}.

Before continuing to the reconstruction of $\vec{x}$, one should note that $\vec{y}$ includes complex values that cannot be kept in a binary format with infinite precision. Accordingly, digital storage and/or transmission of $\vec{y}$ requires lossy (or at least near lossless) compression. 

Let us describe a general lossy compression procedure by the function $ C: \mathbb{R}^N \rightarrow \mathcal{B} $ that maps the $ N $-dimensional signal domain to a discrete set $ \mathcal{B} $ of compressed representations in binary forms of various lengths. The compression of $ \vec{w}\in \mathbb{R}^N $ is denoted by $\textit{b} = C \left( \vec{w} \right)$, where $ \textit{b} \in \mathcal{B} $ is the binary compressed data to store or transmit. 
The decompression of $\textit{b}$ is done via $\vec{v} = F \left( \textit{b} \right)$, 
where $ F: \mathcal{B} \rightarrow \mathcal{S} $ maps binary compressed representations from $ \mathcal{B} $ to their respective decompressed signals in the discrete set $ \mathcal{S} \subset \mathbb{R}^N $. 

In the case of image compression, the decompressed signal $ \vec{v} $ may be displayed. Moreover, in our MRI acquisition setting, the goal is to get a decompressed image $\vec{v}$ that approximates well the unknown visual signal $\vec{x}$, where the approximation quality is constrained by the number of bits utilized for the compression. 

Now we turn to define the optimization problem for the joint task of MRI reconstruction and compression. 
Since $\vec{x}$ is unknown and only its degraded measurement $\vec{y}$ is available, the suggested distortion metric resembles the fidelity term in inverse problem (see also the discussion in \cite{dar2018system}) taking here the form of 
\begin{IEEEeqnarray}{rCl}
	\label{eq:network structure - expected distortion}
	D_{\mtx{A}}\left( \vec{y}, \vec{v} \right) \triangleq \frac{1}{N} \left\| { \vec{y} - \mtx{A} \vec{v} } \right\|_2^2 
\end{IEEEeqnarray}
where the matrix $\mtx{A}$ was defined in the MRI acquisition model in (\ref{eq:MRI acquisition model}). The idea in (\ref{eq:network structure - expected distortion}) is that, after applying the subsampling operator $\mtx{A}$, the decompressed image $\vec{v}$ should be close to the given measurements $\vec{y}$ up to the noise term. While the minimal desired value of $D_{\mtx{A}}\left( \vec{y}, \vec{v} \right)$ should be a positive value that depends on the noise level \cite{dar2018system}, this value is not required for our method. 

Our goal is to optimize the rate-distortion performance of the joint reconstruction and compression of the MRI measurements given in $ \vec{y} $. For this, we formulate the task as   
\begin{IEEEeqnarray}{rCl}
	\label{eq:rate-distortion optimization - Lagrangian}
	\hat{ \vec{v}} = \underset{ \vec{v}\in\mathcal{S} }{\text{argmin}}
	~~ { R \left( \vec{v} \right) + \lambda D_{\mtx{A}}\left( \vec{y}, \vec{v} \right) + \alpha {\rm TV}\left( \vec{v} \right) }
\end{IEEEeqnarray}
where $R \left( \vec{v} \right)$ evaluates the length of the binary compressed description $  \textit{b} \in \mathcal{B} $ matched to the decompressed signal $ \vec{v}\in\mathcal{S} $, $D_{\mtx{A}}\left( \vec{y}, \vec{v} \right)$ is the overall distortion as defined in (\ref{eq:network structure - expected distortion}).
The term ${\rm TV}\left( \vec{v} \right)$ is the total variation of the decompressed image, which is defined in the discrete setting as
\begin{equation*}
\operatorname{TV}(\vec{v}) = \sum_{(i,j)\in \Omega} \sqrt{| \nabla_1 v(i,j) |^2 + | \nabla_2 v(i,j) |^2} 
\label{eq:dtv}
\end{equation*} 
where $v$ is the two-dimensional organization of the vector $\vec{v}$ on the discrete two-dimensional image grid $\Omega$.
Then, $\nabla_1$ and $\nabla_2$ are the discrete gradient operators in the horizontal and vertical directions of $\Omega$, respectively.
The parameters $\lambda\ge 0$ and $\alpha\ge 0$ reflect the relative importance of minimizing the distortion and TV regularization, respectively. Moreover, the values of $\lambda$ and $\alpha$ induce the actual number of bits used for the compressed binary representation that is coupled with $\hat{ \vec{v}}$ (such coding without an explicitly specified bit-rate constraint is common, e.g., in video coding \cite{sullivan1998rate,sullivan2012overview}) 

The unconstrained Lagrangian optimization form in (\ref{eq:rate-distortion optimization - Lagrangian}) resembles the contemporary compression formulations via rate-distortion optimizations (see, e.g., \cite{ortega1998rate,sullivan1998rate,sullivan2012overview}). Yet, importantly note that in our case we do not only optimize the compression rate and distortion but also include an additional term for (total variation) regularization of the decompressed result. This is a significant feature that regularizes the decompression result already at the compression stage and not at the post decompression stage as usually done in compression-artifact reduction methods (e.g., \cite{dar2016postprocessing}). Also note that our proposed method resolves compression artifacts together with the degradation originating in the MRI acquisition.

We consider compression of high-dimensional signals (namely, $ N $ is large), hence, the discrete set $ \mathcal{S} $ is extremely large such that a direct solution of the discrete optimization form in (\ref{eq:rate-distortion optimization - Lagrangian}) is impractical for non-trivial forms of $ \mtx{A} $ and for $\alpha>0$. 
In contrast, for $ \mtx{A} = \mtx{I} $ and $\alpha=0$, the optimization in (\ref{eq:rate-distortion optimization - Lagrangian}) reduces to a standard compression form \cite{ortega1998rate,sullivan1998rate} without any reconstruction or regularization aspects and, therefore, can be practically solved using block-based designs that decompose the problem to a set of independent optimizations on non-overlapping blocks of sufficiently low dimensions.

\subsection{Modular Optimization Procedure}
\label{subsec:Modular Optimization Procedure}

We address the computationally challenging, discrete optimization problem (\ref{eq:rate-distortion optimization - Lagrangian}) using the alternating direction method of multipliers (ADMM) technique \cite{boyd2011distributed}.
First, we split the optimization variable such that (\ref{eq:rate-distortion optimization - Lagrangian}) becomes 
\begin{IEEEeqnarray}{rCl}
	\label{eq:rate-distortion optimization - variable splitting}
		&& \hat{ \vec{v}} = \underset{ \vec{v}\in\mathcal{S} , {\vec{z}}\in\mathbb{R}^N }{\text{argmin}}
		~~ { R \left( \vec{v} \right) +  \lambda D_{\mtx{A}}\left( \vec{y}, \vec{z} \right) + \alpha {\rm TV}\left( \vec{z} \right) } 
		\nonumber \\
	    && \text{subject to} ~~  \vec{z} = \vec{v}
\end{IEEEeqnarray}
where $ \vec{z} \in \mathbb{R}^N $ is an auxiliary variable that is not constrained to the discrete set $ \mathcal{S} $.
Then, the scaled form of the augmented Lagrangian and the method of multipliers \cite[Ch. 2]{boyd2011distributed} translate (\ref{eq:rate-distortion optimization - variable splitting}) into the iterative process 
\begin{IEEEeqnarray}{rCl}
	\label{eq:rate-distortion optimization - augmented Lagrangian}
&&  \left\{{ \hat{ \vec{v}}^{(t)}, \hat{\vec{z}}^{(t)} }\right\} = \mathop {{\text{argmin}}}\limits_{ \vec{v}\in\mathcal{S} , {\vec{z}}\in\mathbb{R}^N }  \Bigg\lbrace R \left( \vec{v} \right) +   \lambda D_{\mtx{A}}\left( \vec{y}, \vec{z} \right)
\nonumber \\ 
&& \qquad\qquad\qquad
 + \alpha {\rm TV}\left( \vec{z} \right) + \frac{\beta}{2}{\left\| {  \vec{v} - \vec{z}} +  \vec{u}^{(t)} \right\|_2^2}  \Bigg\rbrace ~~~~
	\\ 
	&&  \vec{u}^{(t+1)} = \vec{u}^{(t)} + \left( \hat{ \vec{v}}^{(t)} - \hat{\vec{z}}^{(t)} \right),
\end{IEEEeqnarray}
where $ t $ is the iteration index, $\vec{u}^{(t)} \in \mathbb{R}^N$ is the scaled dual variable, and $ \beta $ is an auxiliary parameter induced by the augmented Lagrangian.
Then, applying one iteration of alternating minimization on (\ref{eq:rate-distortion optimization - augmented Lagrangian}) provides the ADMM form of the problem that includes a sequence of simpler optimizations 
\begin{IEEEeqnarray}{rCl}
	\label{eq:rate-distortion optimization - ADMM - compression}
	&& \hat{\vec{v}}^{(t)} = \mathop {{\text{argmin}}}\limits_{ \vec{v}\in\mathcal{S} }  R \left( \vec{v} \right) + \frac{\beta}{2}{\left\| {  \vec{v} - \tilde{ \vec{z}}^{(t)} } \right\|_2^2}
	\\
	\label{eq:rate-distortion optimization - ADMM - deconvolution}
	&& \hat{ \vec{z}}^{(t)} = \mathop {\text{argmin}}\limits_{{\vec{z}}\in\mathbb{R}^N }  \lambda D_{\mtx{A}}\left( \vec{y}, \vec{z} \right) + \alpha {\rm TV}\left( \vec{z} \right) + \frac{\beta}{2}{\left\| {  \vec{z} - \tilde{ \vec{v}}^{(t)} } \right\|_2^2} ~~~~~~
	\\
	\label{eq:rate-distortion optimization - ADMM - u update}
	&& \vec{u}^{(t+1)} = \vec{u}^{(t)} + \left( \hat{ \vec{v}}^{(t)} - \hat{\vec{z}}^{(t)} \right)
\end{IEEEeqnarray}
where $ \tilde{ \vec{z}}^{(t)} = \hat{\vec{z}}^{(t-1)} - \vec{u}^{(t)} $ and $ \tilde{ \vec{v}}^{(t)} = \hat{\vec{v}}^{(t)} + \vec{u}^{(t)} $. 
Importantly, the ADMM form decoupled the compression architecture $ \left\lbrace \mathcal{S}, R \right\rbrace $ from the acquisition model and the total variation regularizer. 

The optimization task in (\ref{eq:rate-distortion optimization - ADMM - compression}) corresponds to the Lagrangian rate-distortion optimization form as in standard compression problems with a squared-error distortion metric.
Therefore, similarly to \cite{dar2018optimized,dar2018system,dar2018compression,dar2018restoration}, we replace the solution of (\ref{eq:rate-distortion optimization - ADMM - compression}) with an application of a standard compression (and decompression) method, which does not have to exactly solve the Lagrangian optimization in (\ref{eq:rate-distortion optimization - ADMM - compression}).
The standard compression and decompression are denoted here as 
\begin{IEEEeqnarray}{rCl}
	\label{eq:rate-distortion optimization - ADMM - compression - standard compression}
	{\textit{b}}^{(t)} = {\rm StandardCompress} \left( \tilde{ \vec{z}}^{(t)}, \theta \right)
	\\
	\label{eq:rate-distortion optimization - ADMM - compression - standard decompression}
	\hat{\vec{v}}^{(t)} = {\rm StandardDecompress} \left( {\textit{b}}^{(t)} \right)		
\end{IEEEeqnarray}
where the parameter $ \theta $ generalizes the Lagrange multiplier part in inducing the rate-distortion tradeoff. The proposed procedure (see Algorithm \ref{Algorithm:Proposed Method}) has a generic structure that may consider any compression method together with the MRI reconstruction task. 

The optimization in (\ref{eq:rate-distortion optimization - ADMM - deconvolution}) is over a continuous domain and its cost includes two quadratic terms and a total variation regularizer. 
Hence, (\ref{eq:rate-distortion optimization - ADMM - deconvolution}) can be numerically solved using primal-dual algorithms \cite{chambolle2011first,chambolle2016introduction,esser2010general}, e.g., see  \cite{corona2019enhancing}.  
Since each of the three terms in (\ref{eq:rate-distortion optimization - ADMM - deconvolution}) has its own parameter, we can degenerate one of the parameters (e.g., $\lambda$) as can be observed in Algorithm \ref{Algorithm:Proposed Method}.

\begin{algorithm}
	\caption{Joint Optimization of MRI Reconstruction and Compression with Total Variation Regularization}
	\label{Algorithm:Proposed Method}
	\begin{algorithmic}[1]
		\State Inputs: $ \vec{y} $, $ \theta $,  $ \alpha $, $ \beta $.
		\State  Initialize $t = 0,  {\hat{\vec{z}}}^{(0)} = \vec{x}, \vec{u}^{(1)} = \vec{0}$.
		\Repeat 
		
		\State $ t \gets t + 1$
		
		\State $ \tilde{ \vec{z}}^{(t)} = \hat{\vec{z}}^{(t-1)} - \vec{u}^{(t)} $
		
		\State $ {\textit{b}}^{(t)} = {\rm StandardCompress} \left( \tilde{ \vec{z}}^{(t)}, \theta \right) $
		\State $ \hat{\vec{v}}^{(t)} = {\rm StandardDecompress} \left( {\textit{b}}^{(t)} \right) $
		
		\State $ \tilde{ \vec{v}}^{(t)} = \hat{\vec{v}}^{(t)} + \vec{u}^{(t)} $
		\State \resizebox{.85\hsize}{!}{$ \hat{ \vec{z}}^{(t)} = \mathop {\text{argmin}}\limits_{{\vec{z}}\in\mathbb{R}^N }   D_{\mtx{A}}\left( \vec{y}, \vec{z} \right) + \alpha {\rm TV}\left( \vec{z} \right) + \frac{\beta}{2}{\left\| {  \vec{z} - \tilde{ \vec{v}}^{(t)} } \right\|_2^2} $}
		
		\State $\vec{u}^{(t+1)} = \vec{u}^{(t)} + \left( \hat{ \vec{v}}^{(t)} - \hat{\vec{z}}^{(t)} \right)$
		
		\Until{stopping criterion is satisfied}
		\State Output: The binary compressed data $ {\textit{b}}^{(t)} $.
	\end{algorithmic}
\end{algorithm}

\section{Experimental Results}
\label{sec:Experimental Results}

In this section we demonstrate the great potential of our approach for the joint optimization of MRI reconstruction and compression. 

We evaluate our reconstruction quality using the Peak Signal-to-Noise Ratio (PSNR) defined as
\begin{equation*}
    {\rm PSNR} = 10 \log_{10} \left( \frac{P^2}{ \frac{1}{N}\|\vec{x} - \hat{\vec{v}}\|_2^2}\right),
\end{equation*}
where $\vec{x}$ is the groundtruth image, $\hat{\vec{v}}$ is the decompressed MRI reconstruction, $P$ is the maximal value attainable by an image pixel and $N$ is the number of pixels in the image. 

We evaluate the performance of our proposed approach using the state-of-the-art image (and video) compression standard HEVC \cite{sullivan2012overview} in its BPG implementation \cite{hevc_software_bpg}.  We consider different compression levels by setting quantization parameter (QP) values between 4 to 49 in jumps
of 3. 
In the implementation of our proposed approach we empirically set $\beta$ to a value depending on the specific quantization parameter (QP) given to the HEVC compression, specifically, 
 \begin{equation*}
     \beta = 5.5 - 0.1\text{QP} 
 \end{equation*}
where QP values are integers between 0 and 51 (where a lower QP value corresponds to a higher reconstruction quality). The stopping criterion in our experiments was the arrival to a maximal number of 40 iterations or when $\hat{\vec{v}}^{(t)}$ and $\hat{\vec{z}}^{(t)}$ are detected to converge or diverge. The convergence/divergence is defined based on the total absolute difference between $\hat{\vec{v}}^{(t)}$ and $\hat{\vec{z}}^{(t)}$ in each iteration, i.e., $\hat{w}^{(t)}\triangleq \|\hat{\vec{v}}^{(t)} - \hat{\vec{z}}^{(t)}\|_{1}$. Specifically, we determine convergence when $|w^{(t)} - w^{(t-1)}| < 0.5$ for three consecutive iterations, whereas divergence is identified when
$|w^{(t)} - w^{(t-1)}|> 50$.

The datasets used in our experiments are the following:
\begin{itemize}
    \item Brain\footnote{\url{https://github.com/veronicacorona/multicontrastSegmentation}}: the dataset is taken from \cite{CoronaHBM}. The data is a T$_1$-weighted acquired using a 3D MPRAGE sequence with the following scan parameters: Inversion time (TI) = 1,100 ms, flip angle $\alpha=7^{\circ}$, echo time (TE) = 4.37 ms, receiver bandwidth (RB) = 140 Hz per pixel, echospacing = 11.1 ms, repetition time (TR) = 2,500 ms; $256 \times 256 \times 192$ matrix dimensions. 
    We refer to these images as Brain 1, 2 and 3. 
    \item Liver\footnote{\url{http://www.vision.ee.ethz.ch/~organmot/chapter_download.shtml}}: The datasets are 4DMRI data acquired during free-breathing of the right liver lobe \cite{Siebenthal2007}. It was acquired on a 1.5T Philips Achieva system using a T$_1$-weighted gradient echo sequence, TR=3.1 ms, slices=25,  matrix size = $195\times166$, over roughly one hour on 22 to 30 sagittal slices and a temporal resolution of $2.6-2.8$ Hz.
\end{itemize}

\subsection{Results for Settings Without Total-Variation Regularization}

\begin{figure*}
    \centering
    \subfloat[]{
   \raisebox{1.5cm}{\rotatebox[origin=t]{90}{\colorbox{gray!10}{Without TV Reg.}}}}
\subfloat[(a) Brain 1]{\includegraphics[width=0.22\textwidth]{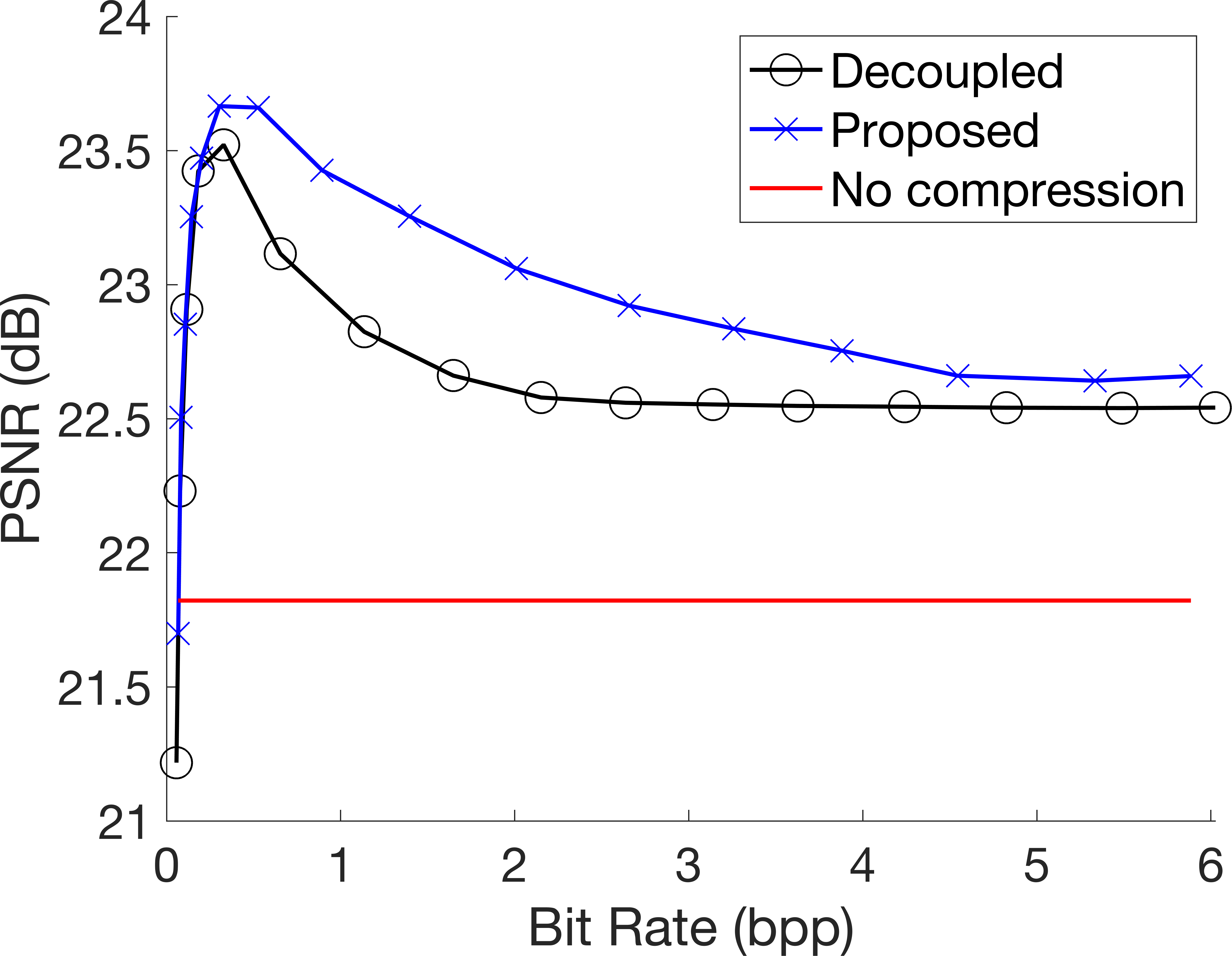}}~
\subfloat[(b) Brain 2]{\includegraphics[width=0.22\textwidth]{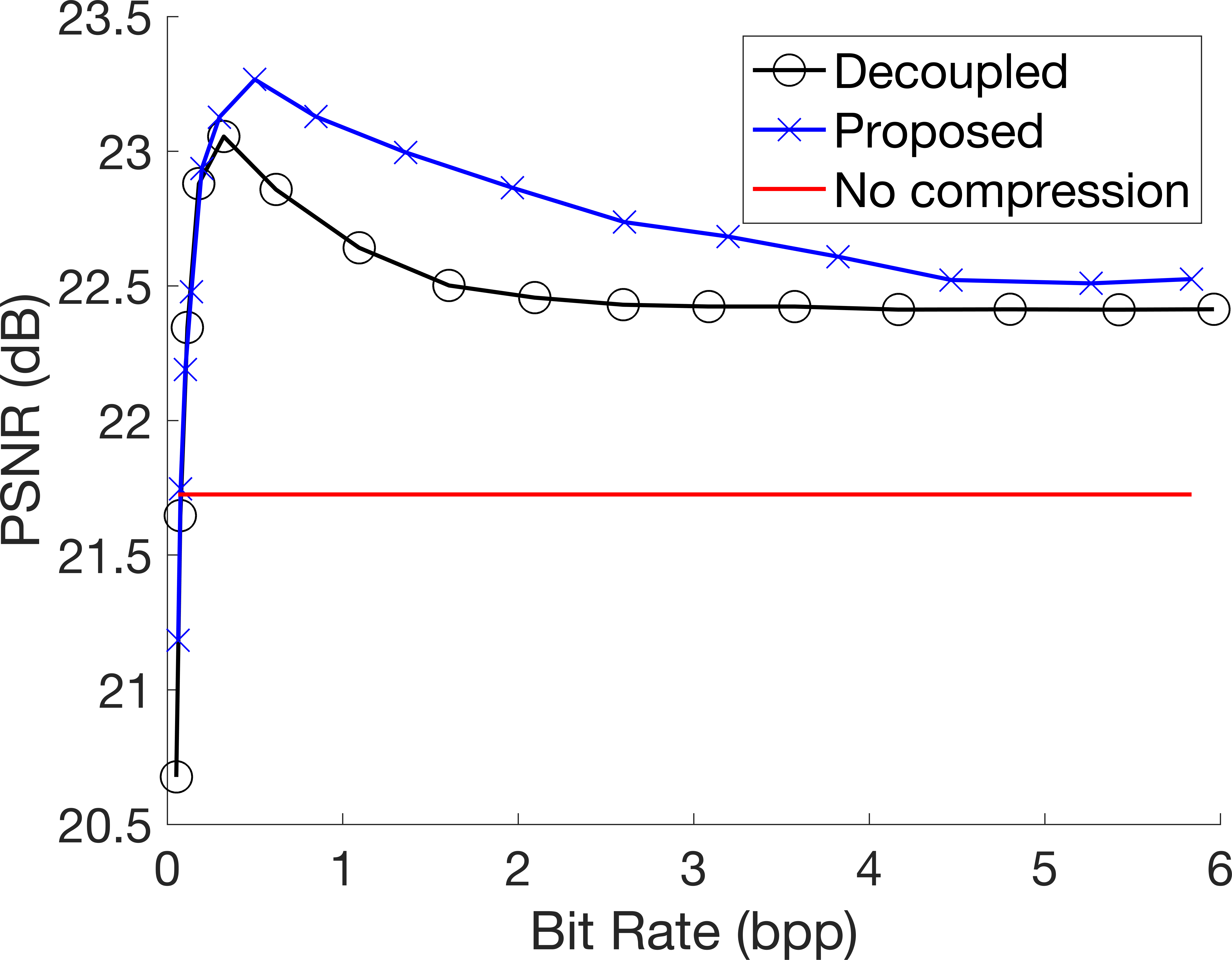}}~
\subfloat[(c) Brain 3 ]{\includegraphics[width=0.22\textwidth,]{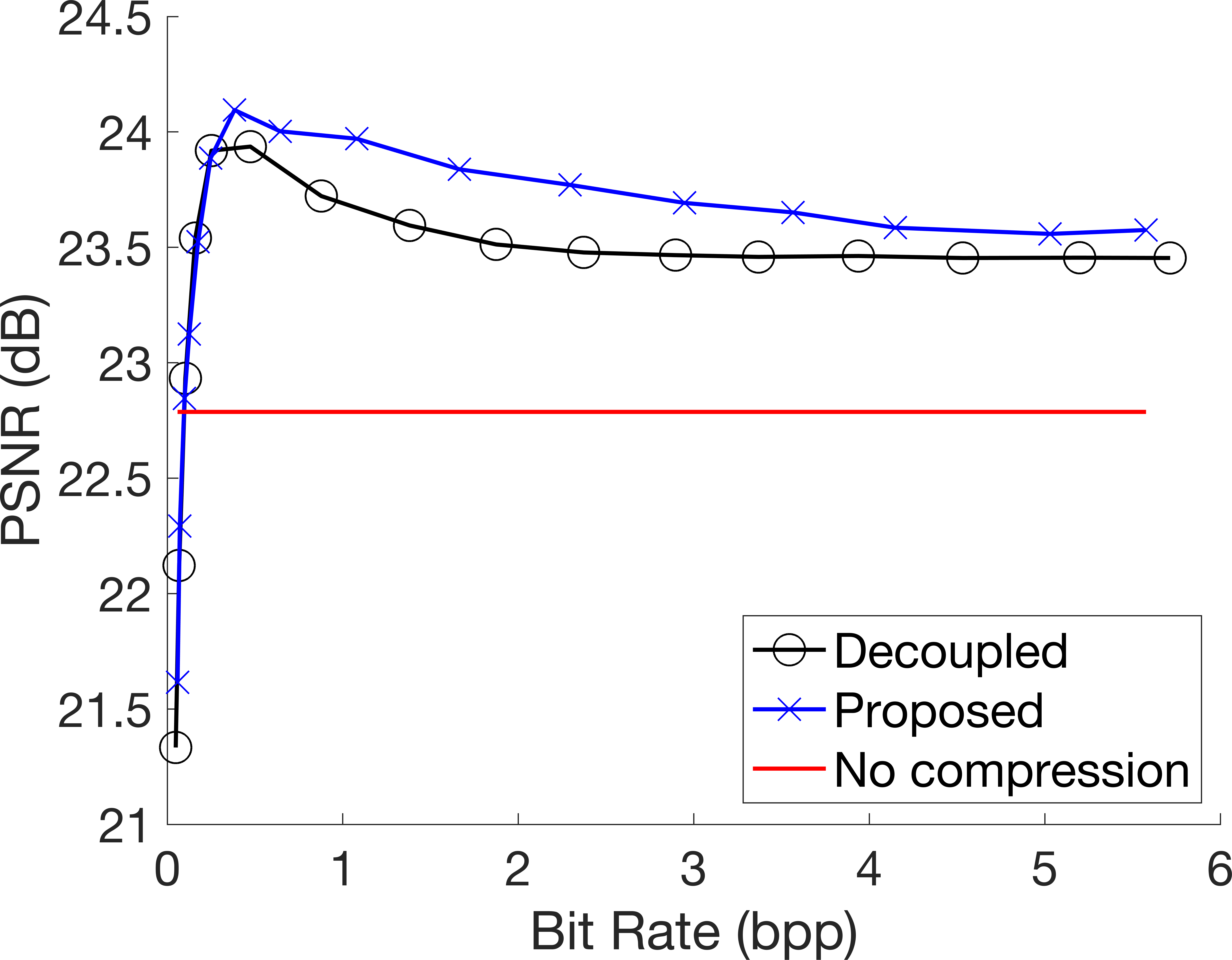}}~
\subfloat[(d) Liver]{\includegraphics[width=0.22\textwidth,]{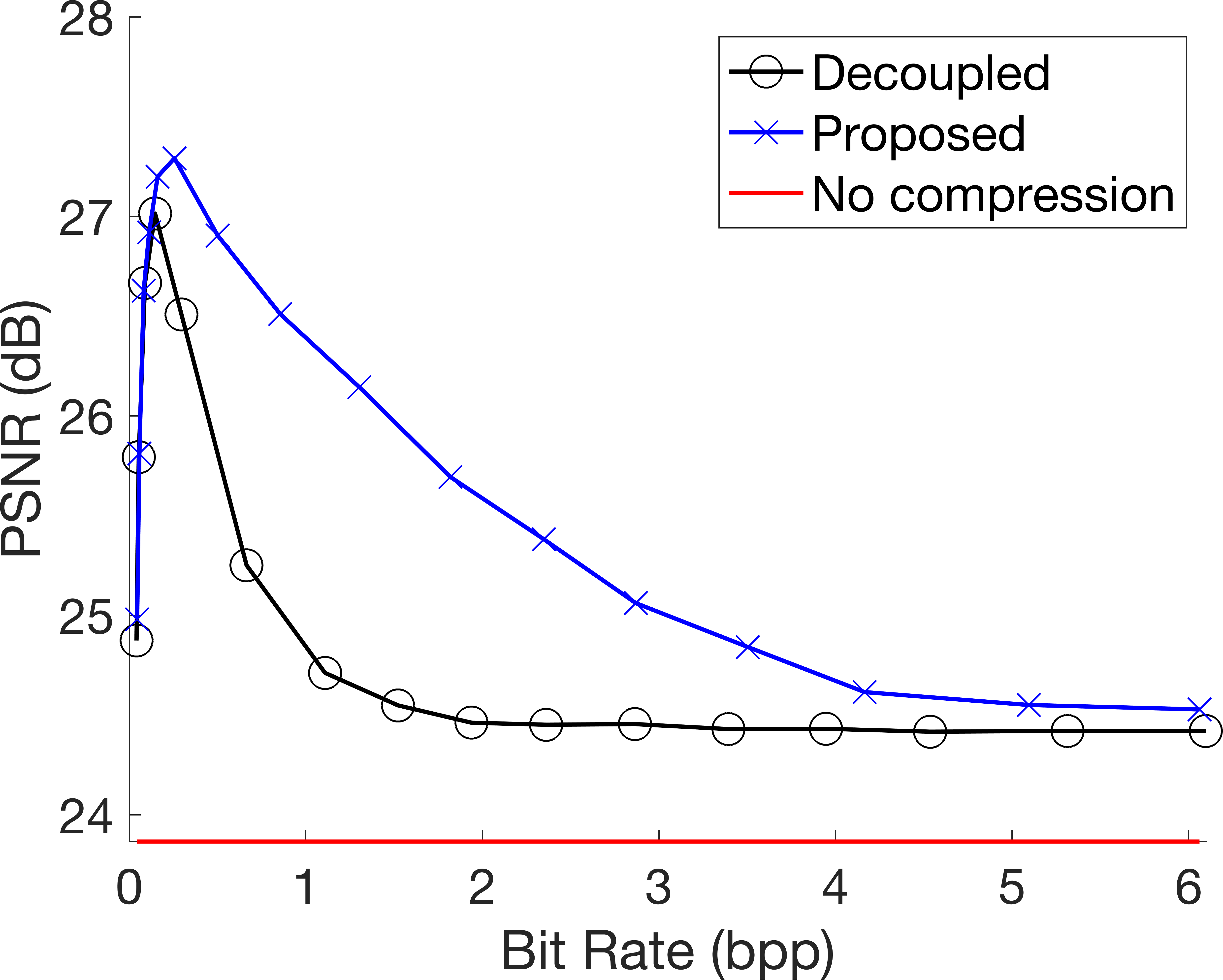}}
\caption{PSNR-bitrate curves comparing the proposed joint optimization without TV-regularization (blue lines) to the decoupled MRI reconstruction and compression without TV-regularization (black lines), and to no compression, only zero-filled MRI reconstruction (without TV regularization, red lines). 
All results are for acceleration factor 4x. Each subfigure is for a different dataset. }
\label{fig:all_4x_option1}
\end{figure*}

\begin{figure*}[t]
    \centering
        \subfloat[]{
   \raisebox{1.5cm}{\rotatebox[origin=t]{90}{\colorbox{gray!10}{Without TV Reg.}}}   
}
\subfloat[(a) Full]{\includegraphics[width=0.22\textwidth]{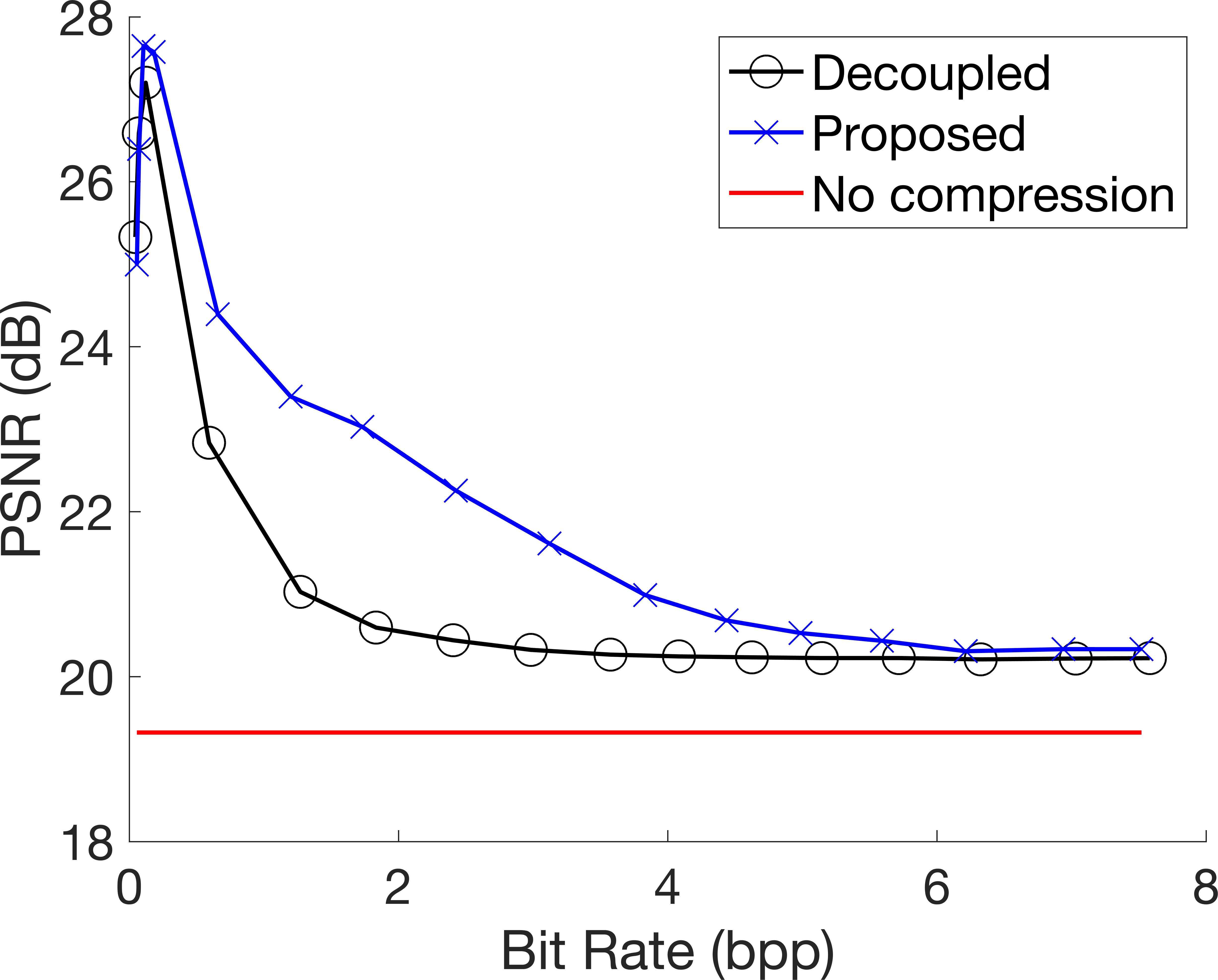}}~
\subfloat[(b) 2x]{\includegraphics[width=0.22\textwidth]{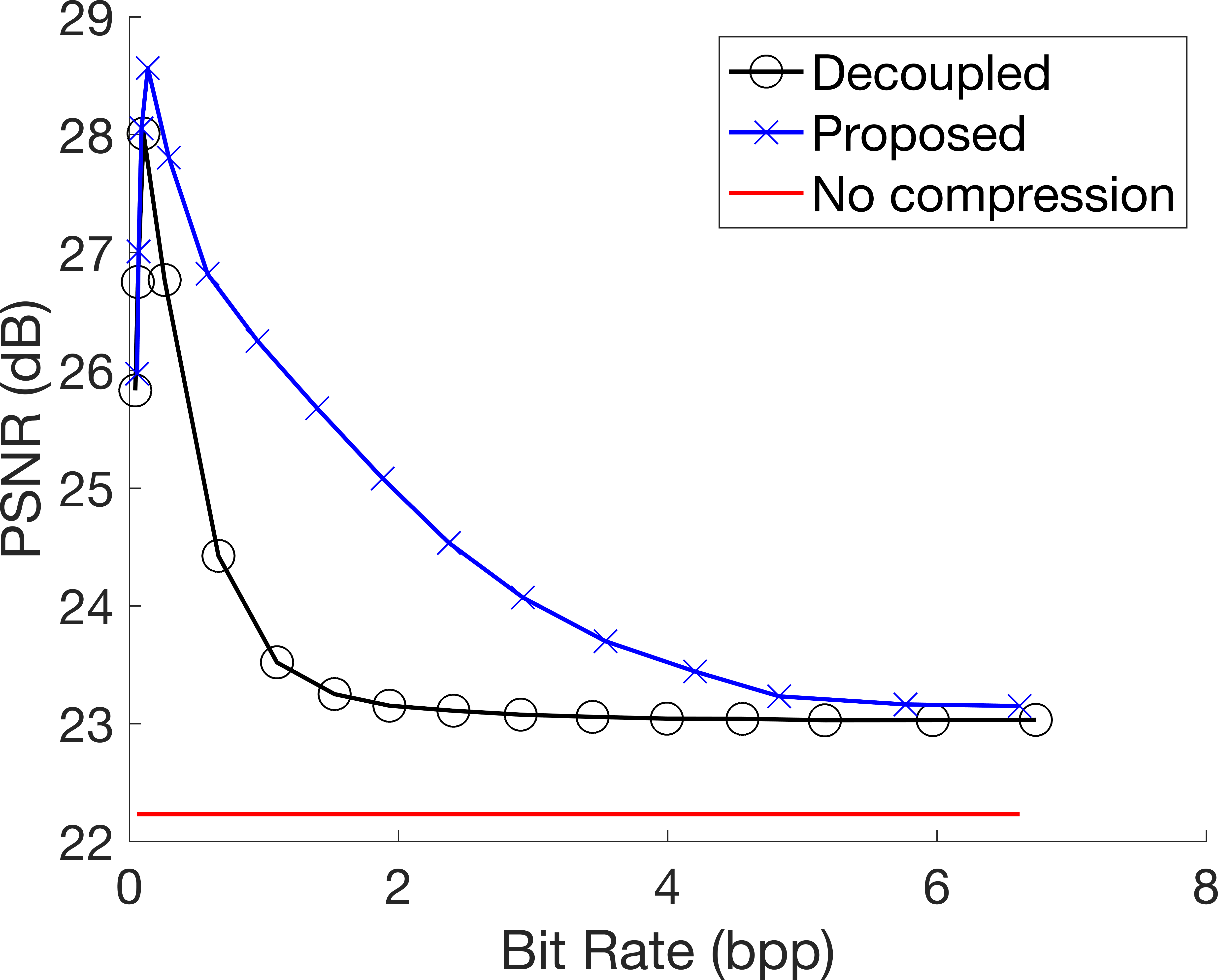}}~
\subfloat[(c) 4x ]{\includegraphics[width=0.22\textwidth,]{figures_psnr/option1_liver_25samp_psnr.png}}~
\subfloat[(d) 8x]{\includegraphics[width=0.22\textwidth,]{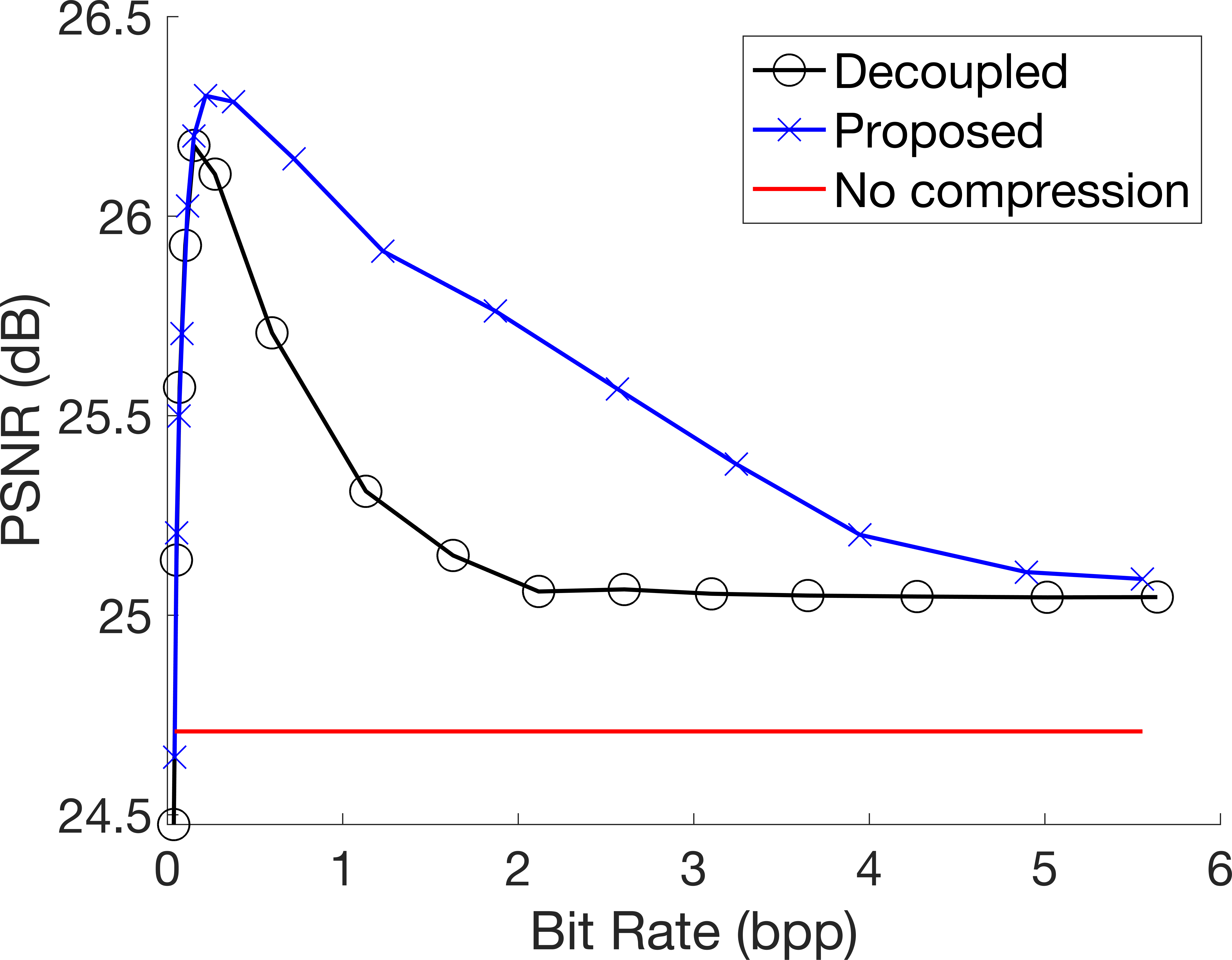}}
\caption{PSNR-bitrate curves comparing the proposed joint optimization without TV-regularization (blue lines) to the decoupled MRI reconstruction and compression without TV-regularization (black lines), and to no compression, only zero-filled MRI reconstruction (without TV regularization) (red lines). 
All results are for the dataset Liver. Each subfigure is for a different acceleration factor.}
\label{fig:option1_liver}
\end{figure*}

\begin{figure*}
    \centering
        \subfloat[]{\raisebox{1.5cm}{\rotatebox[origin=t]{90}{\colorbox{gray!10}{Without TV Reg.}}}   }
\subfloat[(a) Full]{\includegraphics[width=0.22\textwidth]{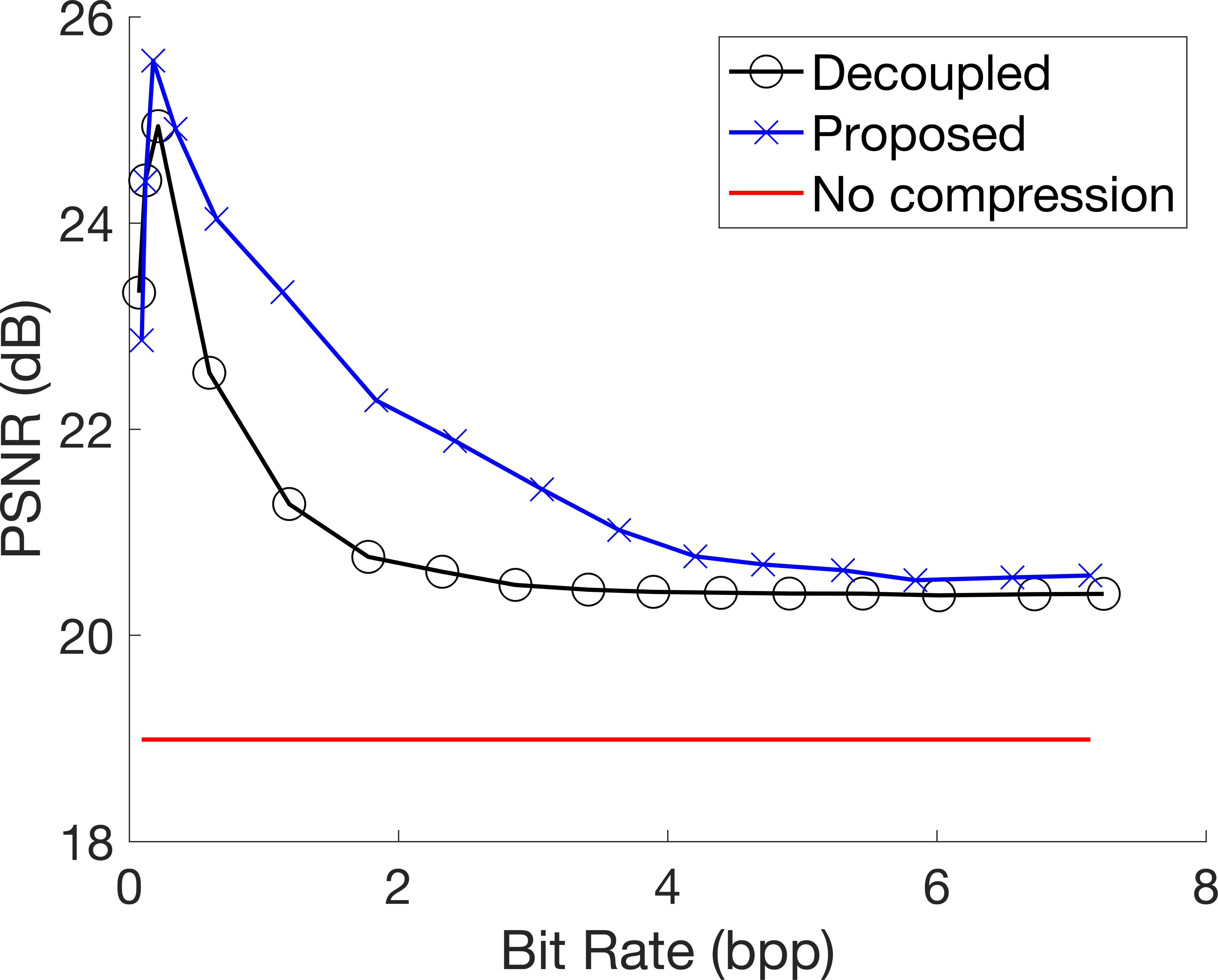}}~
\subfloat[(b) 2x]{\includegraphics[width=0.22\textwidth]{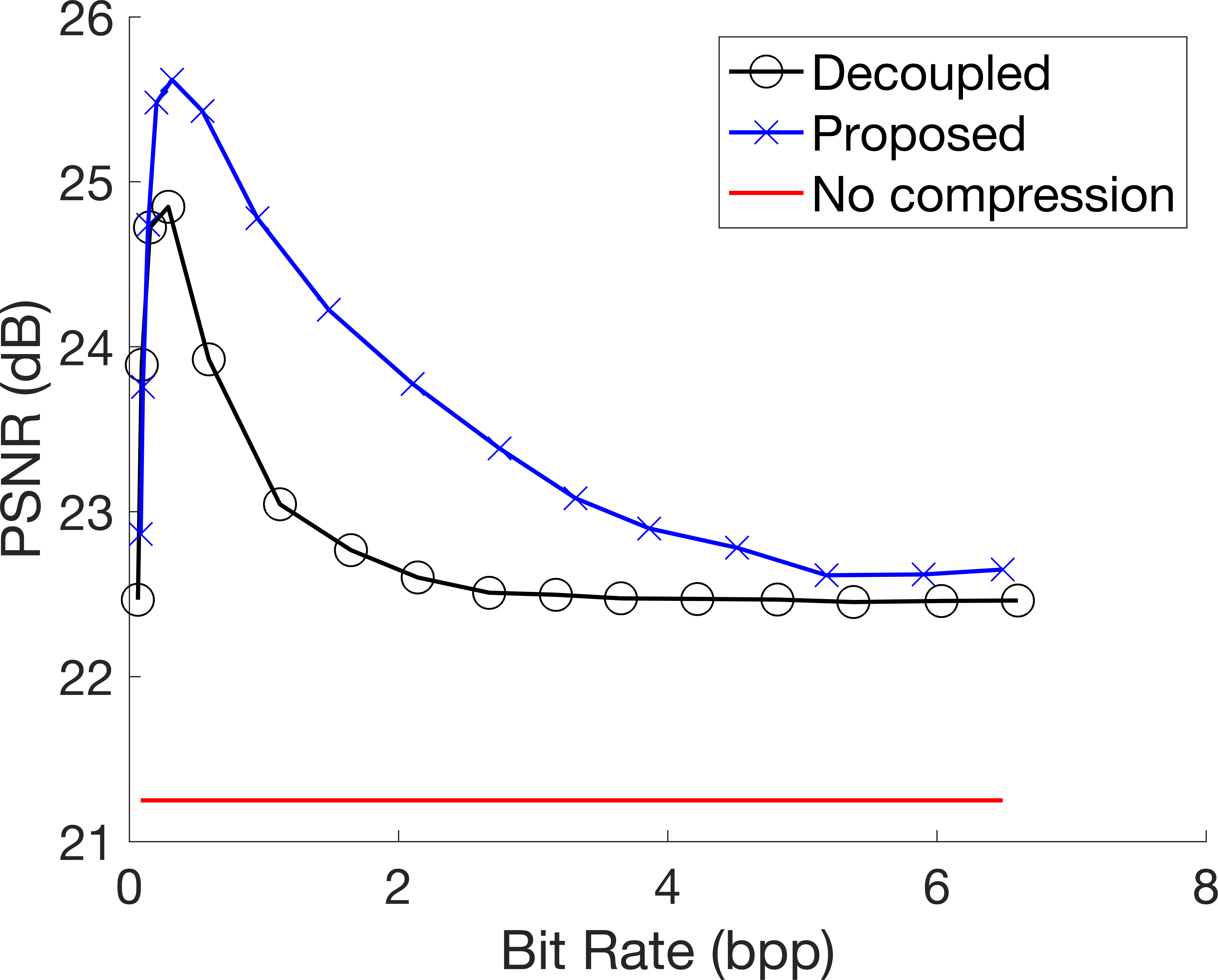}}~
\subfloat[(c) 4x ]{\includegraphics[width=0.22\textwidth,]{figures_psnr/option1_mprage140_25samp_psnr.png}}~
\subfloat[(d) 8x]{\includegraphics[width=0.22\textwidth,]{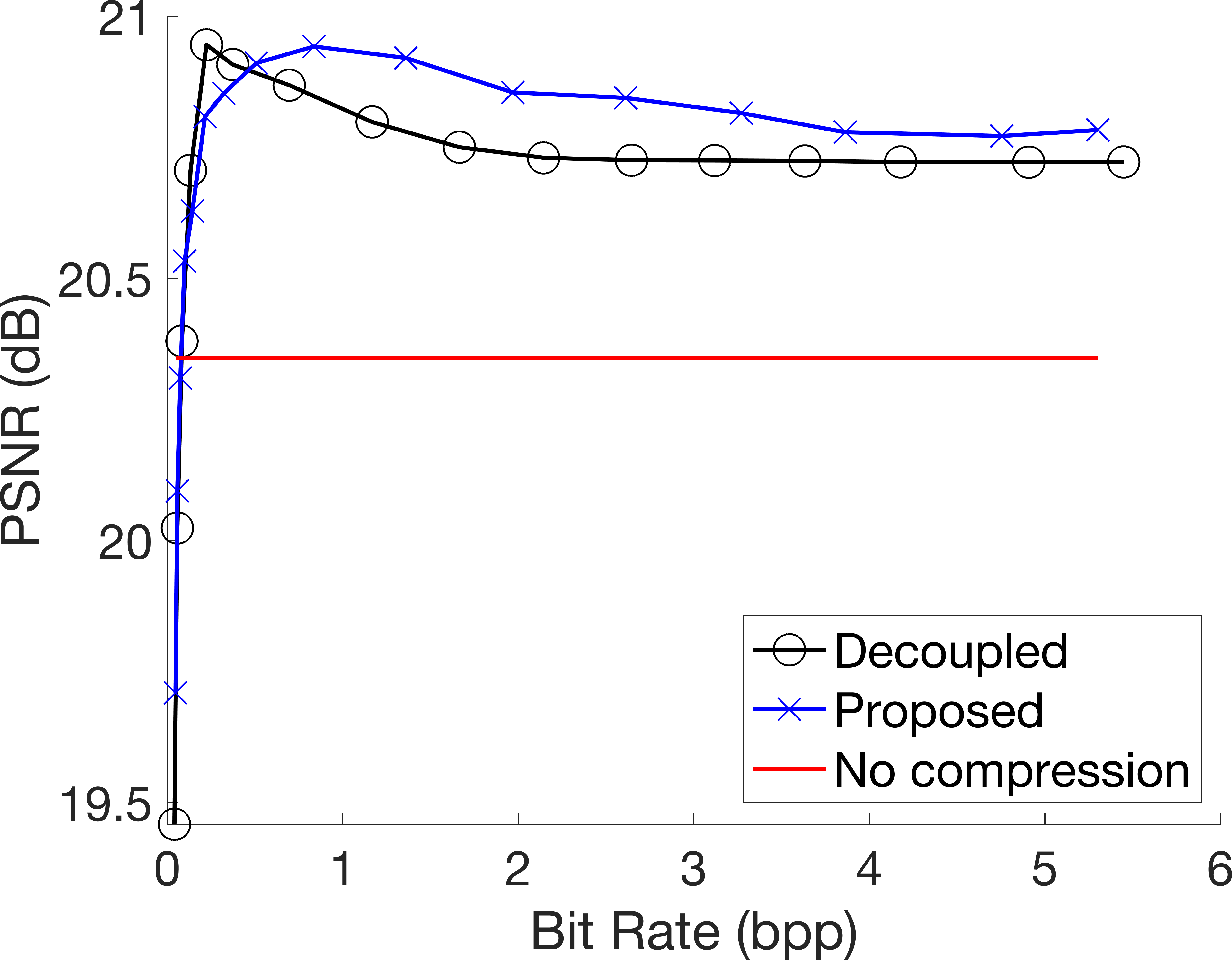}}
\caption{PSNR-bitrate curves comparing the proposed joint optimization without TV-regularization (blue lines) to the decoupled MRI reconstruction and compression without TV-regularization (black lines), and to no compression, only zero-filled MRI reconstruction (without TV regularization) (red lines). 
All results are for the dataset Brain 1. Each subfigure is for a different acceleration factor.}
\label{fig:option1_brain1}
\end{figure*}

Let us start by examining the experimental results for the following methods that do not include total-variation regularization: 
\begin{itemize}
    \item \textit{Proposed joint optimization without TV-regularization}: This approach is obtained by Algorithm \ref{Algorithm:Proposed Method} with $\alpha=0$. This setting will provide joint optimization of the MRI reconstruction and lossy compression such that, effectively, the MRI reconstruction benefits from the implicit regularization introduced by the lossy compression (see \cite{dar2018restoration} for additional details on this type of complexity regularization). 
    The PSNR versus bit-rate results for this setting appear in blue curves in  Figures \ref{fig:all_4x_option1}-\ref{fig:option1_brain1}. 

    \item \textit{Decoupled MRI reconstruction and compression without TV-regularization}: in this competing method, the MR image is reconstructed from the measurements (without regularization), and then the reconstructed MR image goes through a standard compression (see the general description of the flow in Fig.~\ref{fig:decoupled_process}). 
    The unregularized reconstruction stage is done by setting the missing k-space samples to zeros and then computing the inverse Fourier transform of the zero-filled data to get an MR image.
    The PSNR versus bit-rate results for this setting appear in black curves in  Figures \ref{fig:all_4x_option1}-\ref{fig:option1_brain1}. 
    
    \item \textit{No compression, only zero-filled MRI reconstruction (without TV regularization)}: 
    in this setting we evaluate the PSNR of the MR image obtained in a reconstruction process (without TV regularization) operated in 64-bit numerical resolution. In this reconstruction process the missing k-space samples are set to zeros and then inverse Fourier transform is applied to obtain the MR image.  In our comparisons we will use only the PSNR values of this setting that appear as  constant red-lines in Figures \ref{fig:all_4x_option1}-\ref{fig:option1_brain1}.
\end{itemize}


Considering the three options that do not include total-variation regularization, in  Fig.~\ref{fig:all_4x_option1} we show PSNR-bitrate curves for our four datasets and acceleration factor 4x (i.e., keeping 25\% of the coefficients in the Fourier domain subsampling). In these results (in Fig.~\ref{fig:all_4x_option1}) we can note that both the decoupled approach and proposed joint optimization (i.e., Algorithm \ref{Algorithm:Proposed Method} with $\alpha=0$) perform significantly better than the 'no compression' option, suggesting that the lossy compression in our approach acts as a regularizer. 
Moreover, note that our proposed joint optimization further outperforms the decoupled approach, especially in the mid-range of bit rates. 

In Fig.~\ref{fig:option1_liver}-\ref{fig:option1_brain1}, we also report the PSNR-bitrate curves for two datasets (Liver and Brain1) for several acceleration factors. These results demonstrate that our method achieves significant gains also for various acceleration factors.

\begin{figure*}
    \centering
\subfloat[(a) Ground truth]{\includegraphics[width=0.22\textwidth]{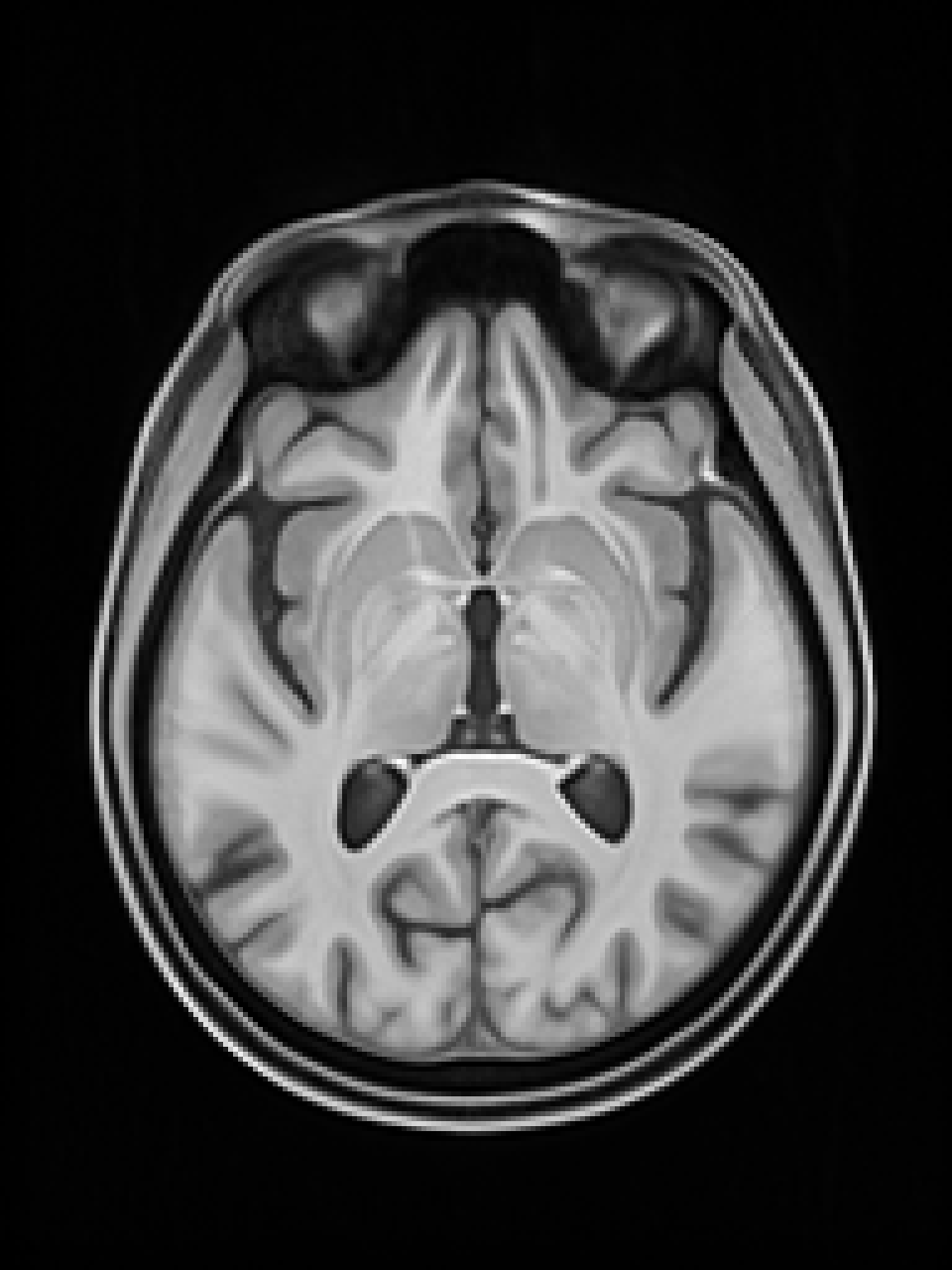}}~
\subfloat[(b) No compression, only zero-filled reconstruction \protect\\ PSNR=21.72]{\includegraphics[width=0.22\textwidth]{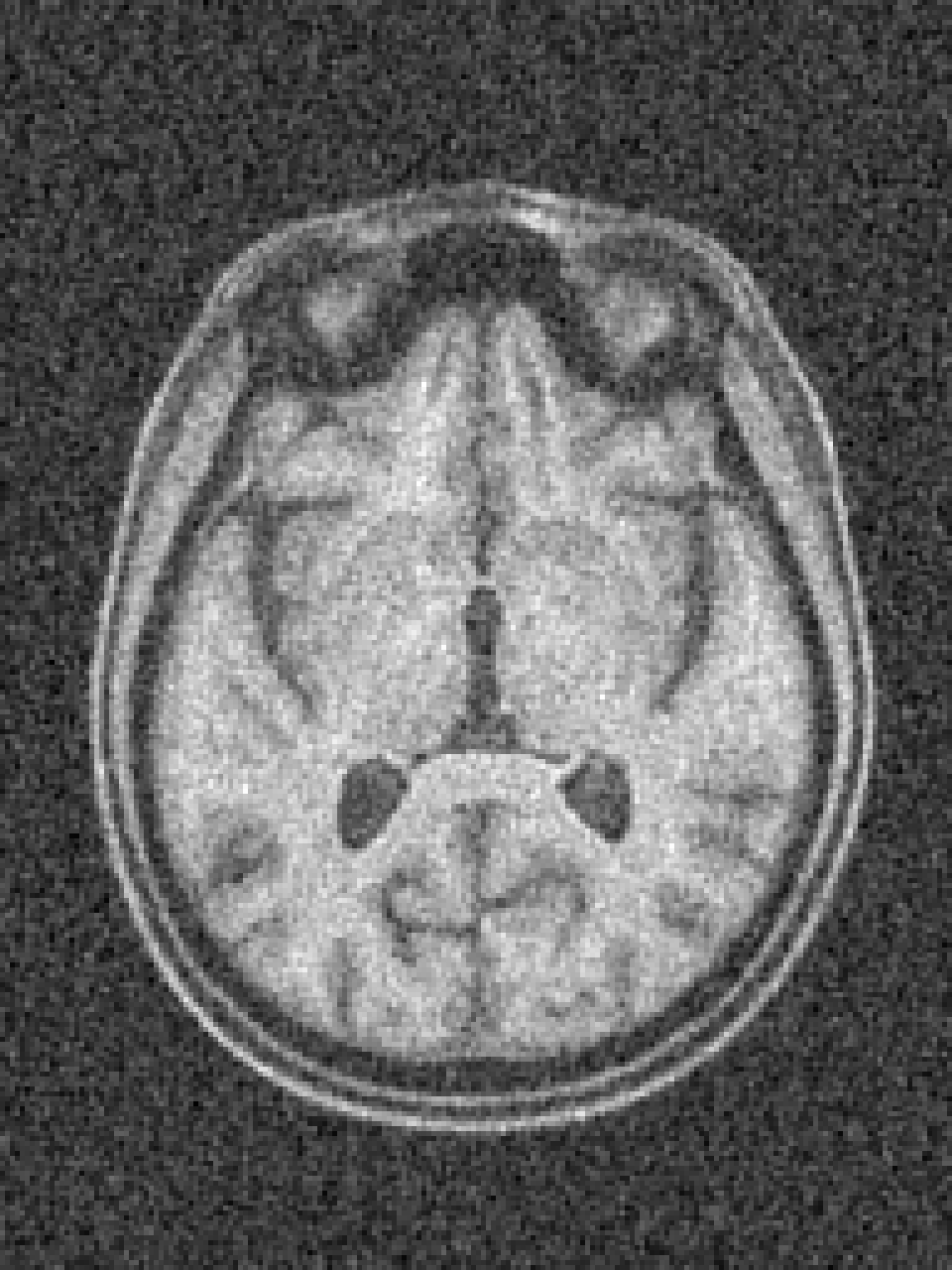}}~
\subfloat[(c) Decoupled without TV reg. \protect\\ PSNR=22.77]{\includegraphics[width=0.22\textwidth,]{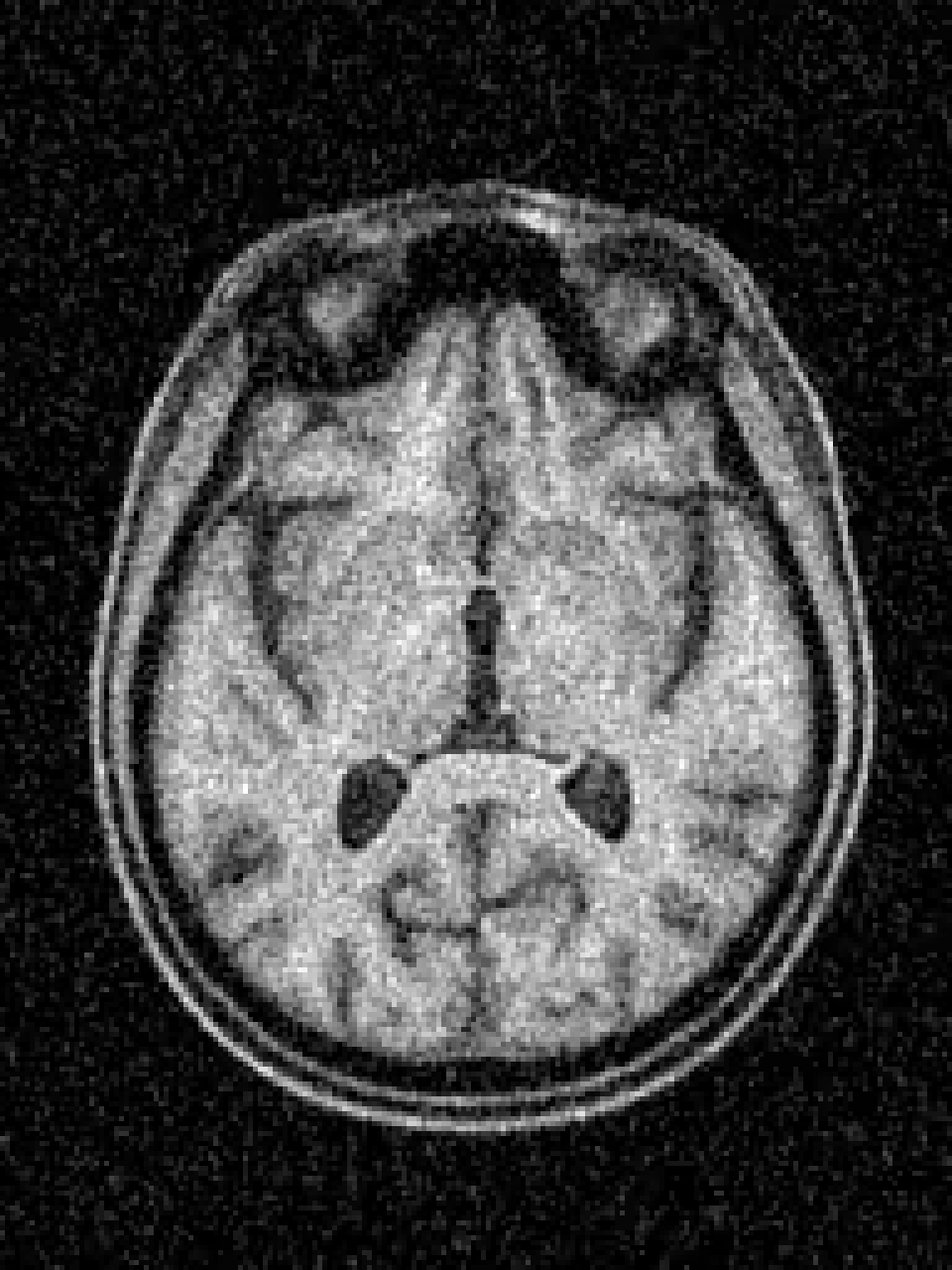}}~
\subfloat[(d) Proposed without TV reg.\protect\\ PSNR=24.78]{\includegraphics[width=0.22\textwidth,]{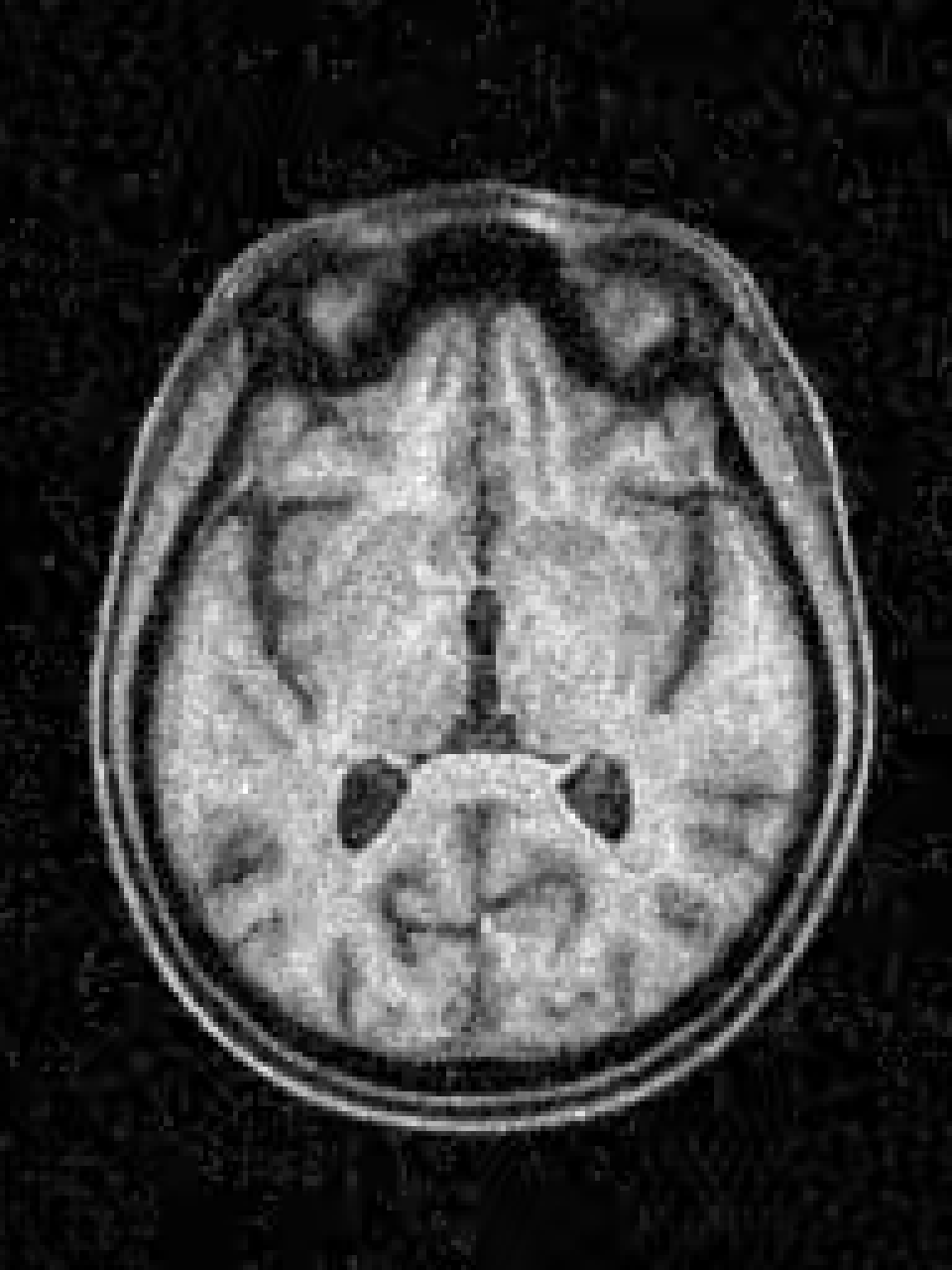}}
\caption{Visual results for data Brain 1 with compression QP 31 and acceleration factor 2x. We show the (a) ground truth image, (b) no-compression zero-filled reconstruction, (c) decoupled MRI reconstruction and compression \textit{without} TV-regularization, and (d) the proposed joint optimization \textit{without} TV-regularization. }
\label{figu:visual_option1_brain1}
\end{figure*}

In \autoref{figu:visual_option1_brain1} we present visual results for the Brain 1 image for the settings without TV regularization. These results clearly show the gains in visual quality obtained using lossy compression in settings that do not include other regularization types (e.g., total variation).

\subsection{Results for Settings With Total-Variation Regularization}

\begin{figure*}[t]
    \centering
            \subfloat[]{\raisebox{1.5cm}{\rotatebox[origin=t]{90}{\colorbox{gray!10}{With TV Reg.}}}   }
\subfloat[(a) Brain 1]{\includegraphics[width=0.22\textwidth]{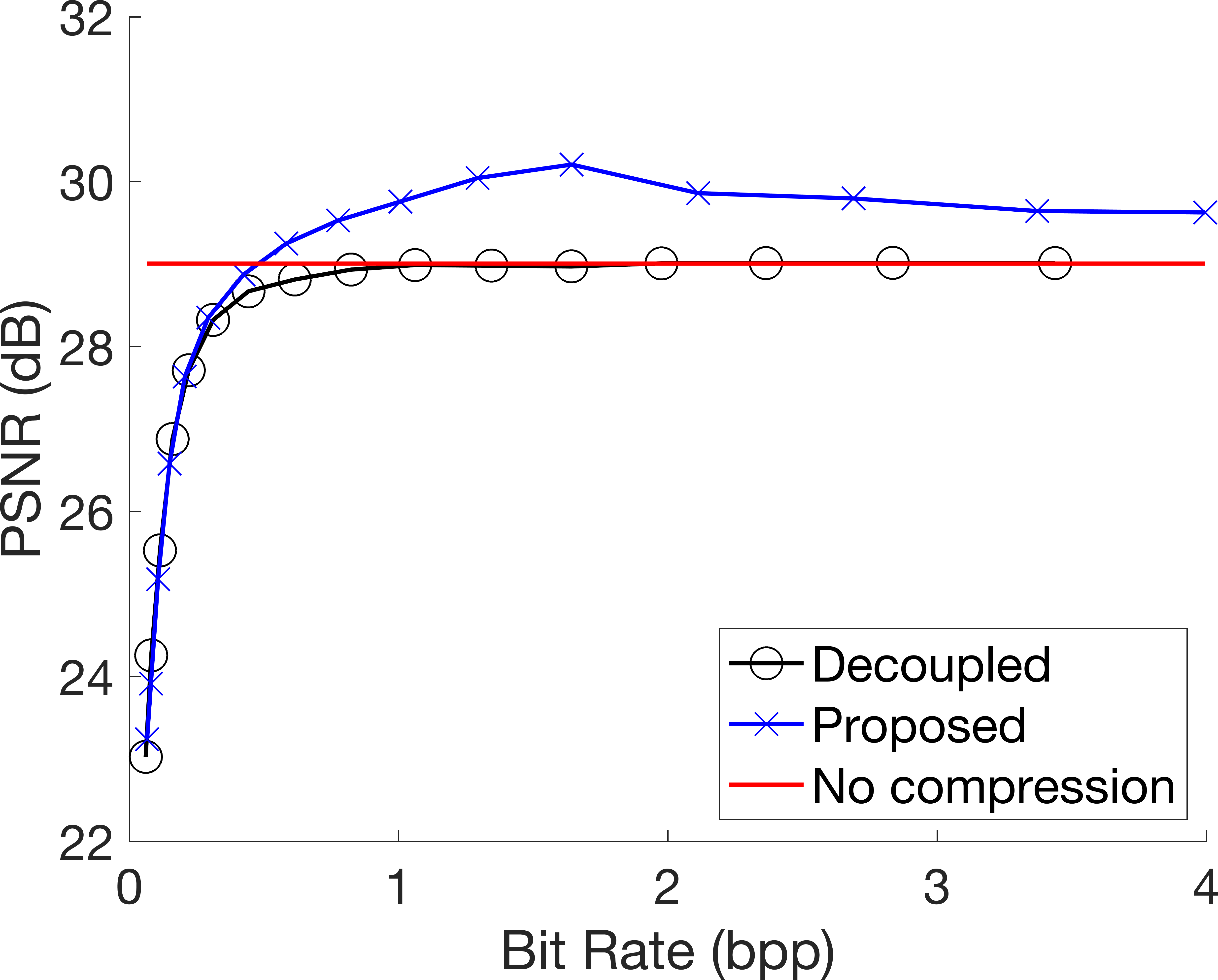}}~
\subfloat[(b) Brain 2]{\includegraphics[width=0.22\textwidth]{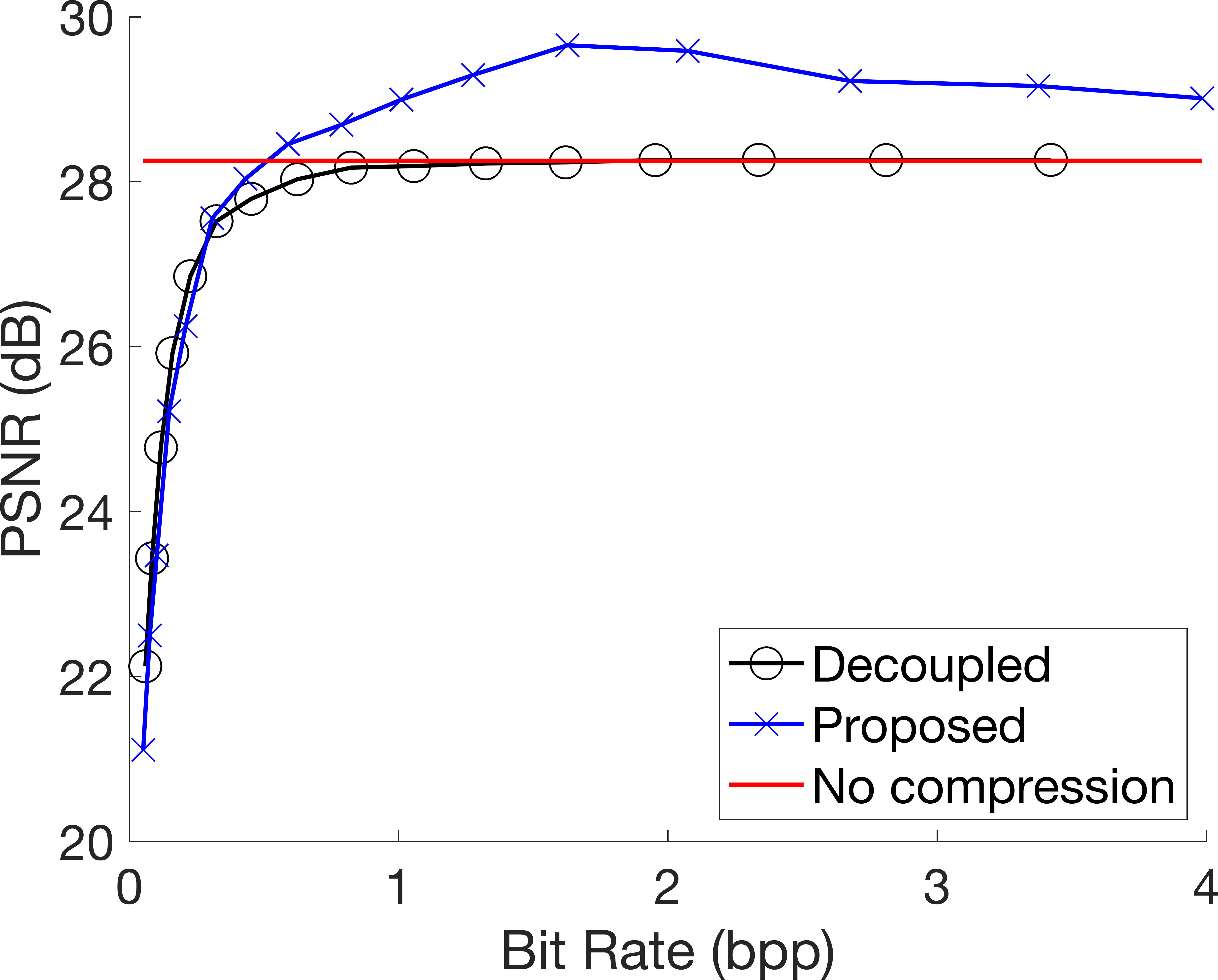}}~
\subfloat[(c) Brain 3 ]{\includegraphics[width=0.22\textwidth,]{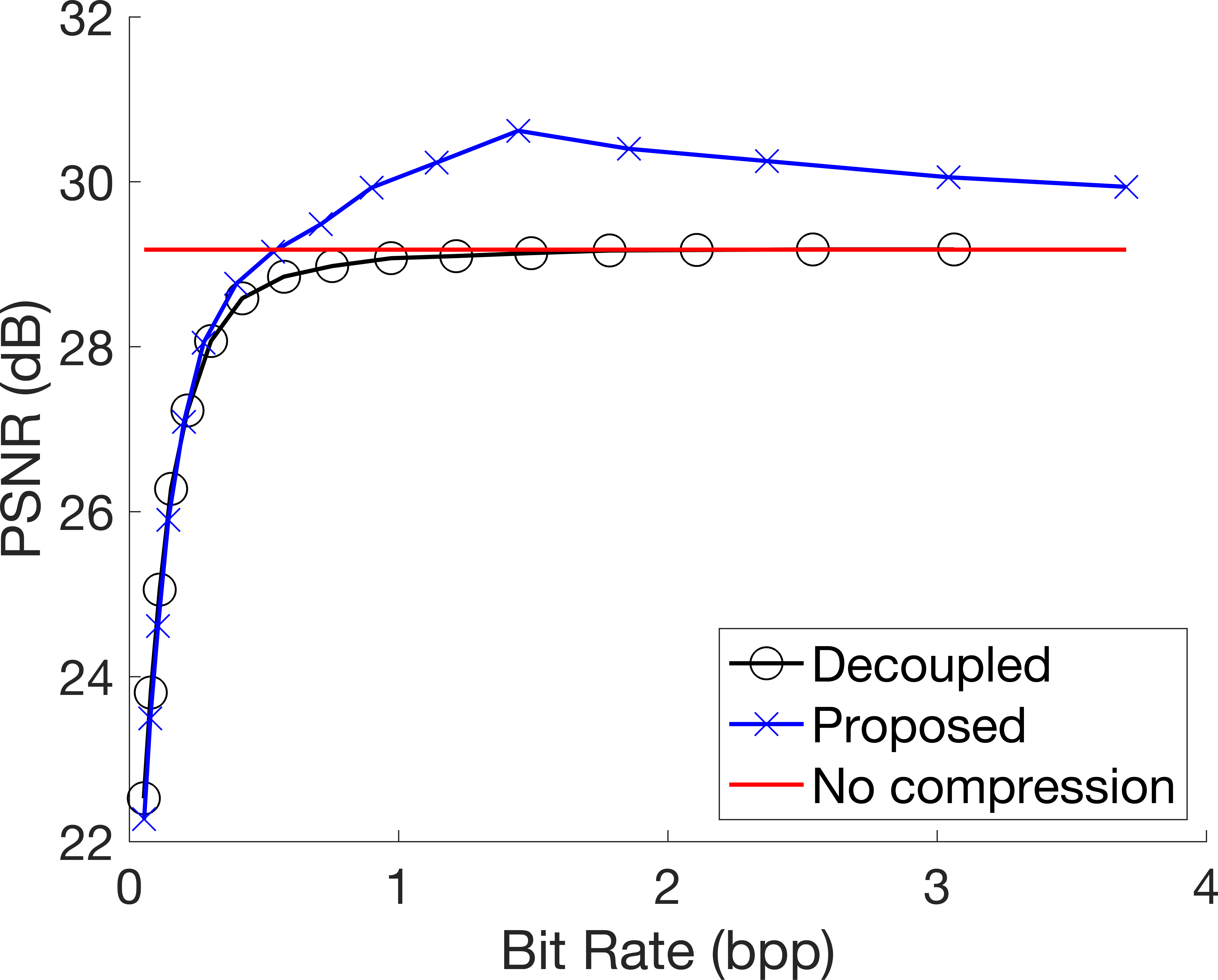}}~
\subfloat[(d) Liver]{\includegraphics[width=0.22\textwidth,]{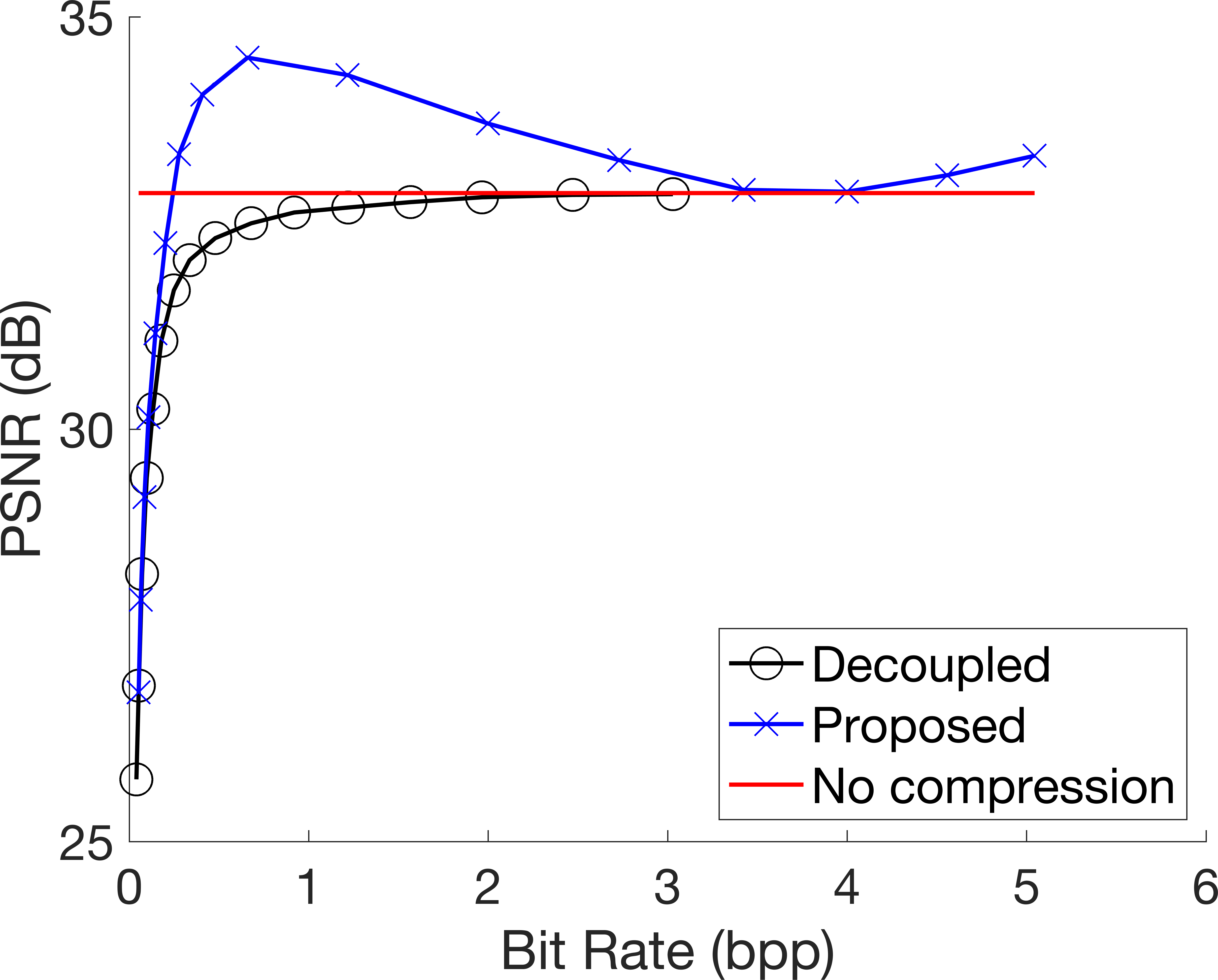}}
\caption{PSNR-bitrate curves comparing the proposed joint optimization using TV-regularization (blue lines) to the decoupled MRI reconstruction and compression using TV-regularization (black lines), and to no compression, only TV-regularized MRI reconstruction (red lines). 
All results are for acceleration factor 4x and regularization parameter $\alpha=0.01$. Each subfigure is for a different dataset. }
\label{fig:all_4x_option2}
\end{figure*}

\begin{figure*}[t!]
    \centering
            \subfloat[]{\raisebox{1.5cm}{\rotatebox[origin=t]{90}{\colorbox{gray!10}{With TV Reg.}}}   }
\subfloat[(a) Full]{\includegraphics[width=0.22\textwidth]{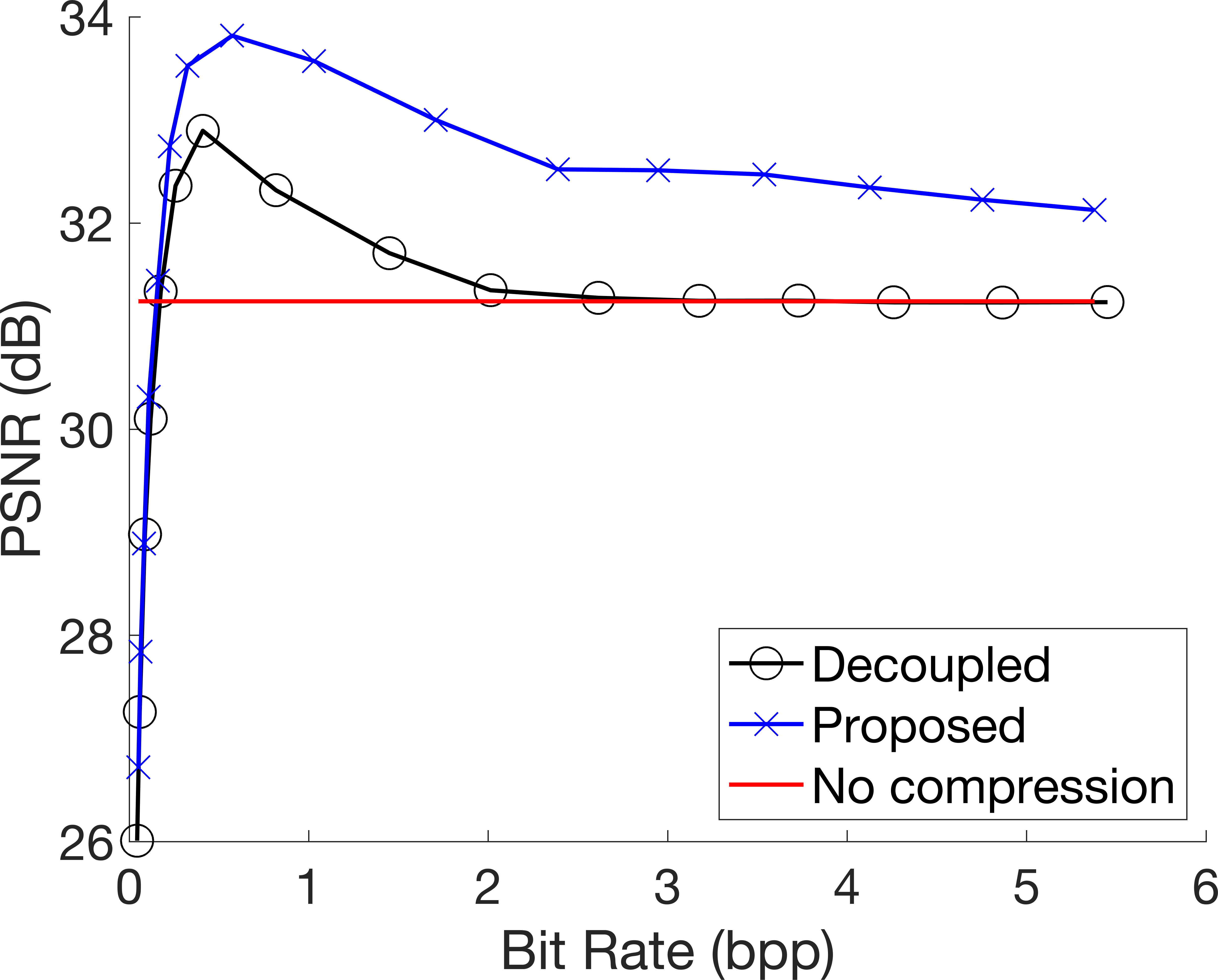}}~
\subfloat[(b) 2x]{\includegraphics[width=0.22\textwidth]{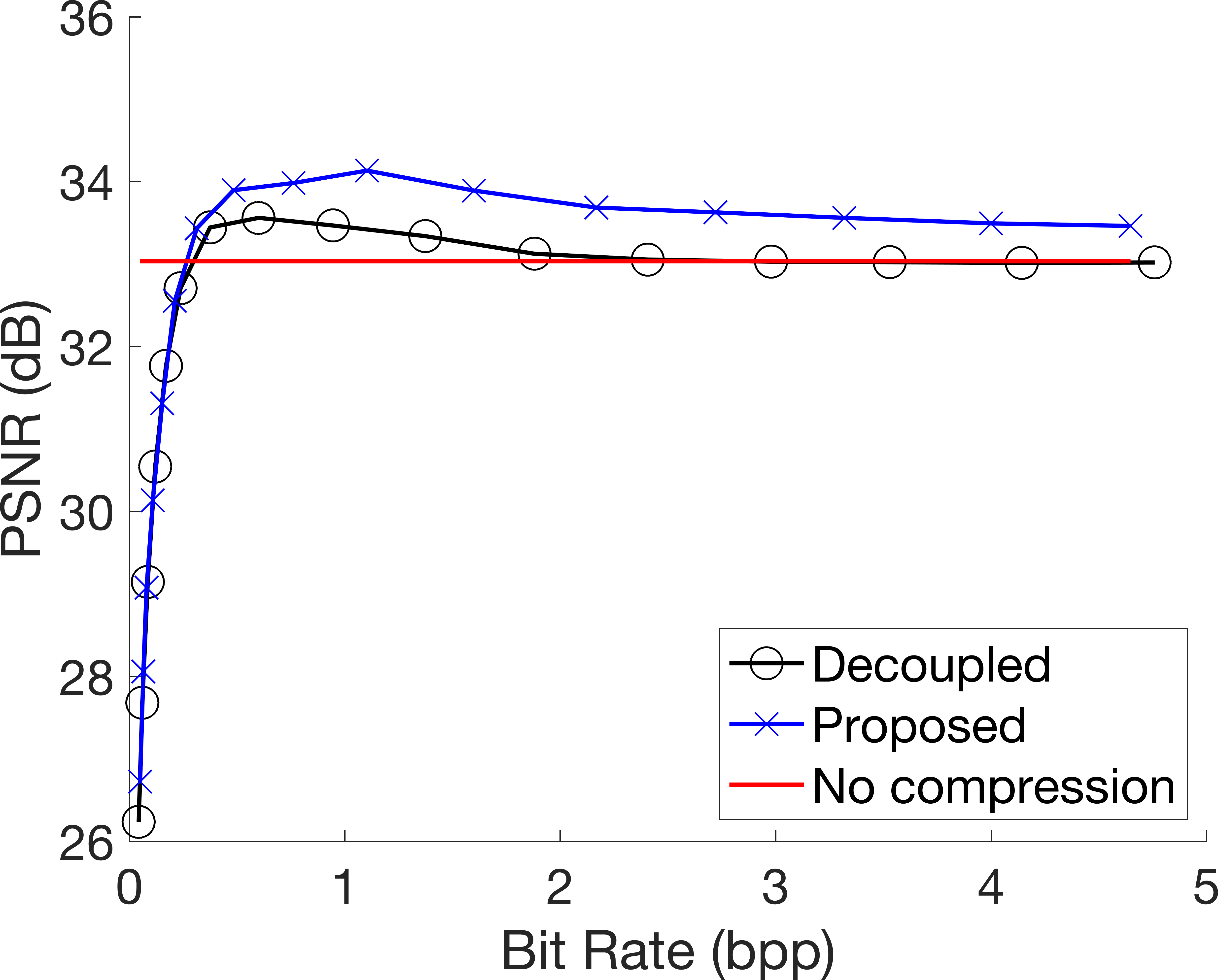}}~
\subfloat[(c) 4x ]{\includegraphics[width=0.22\textwidth,]{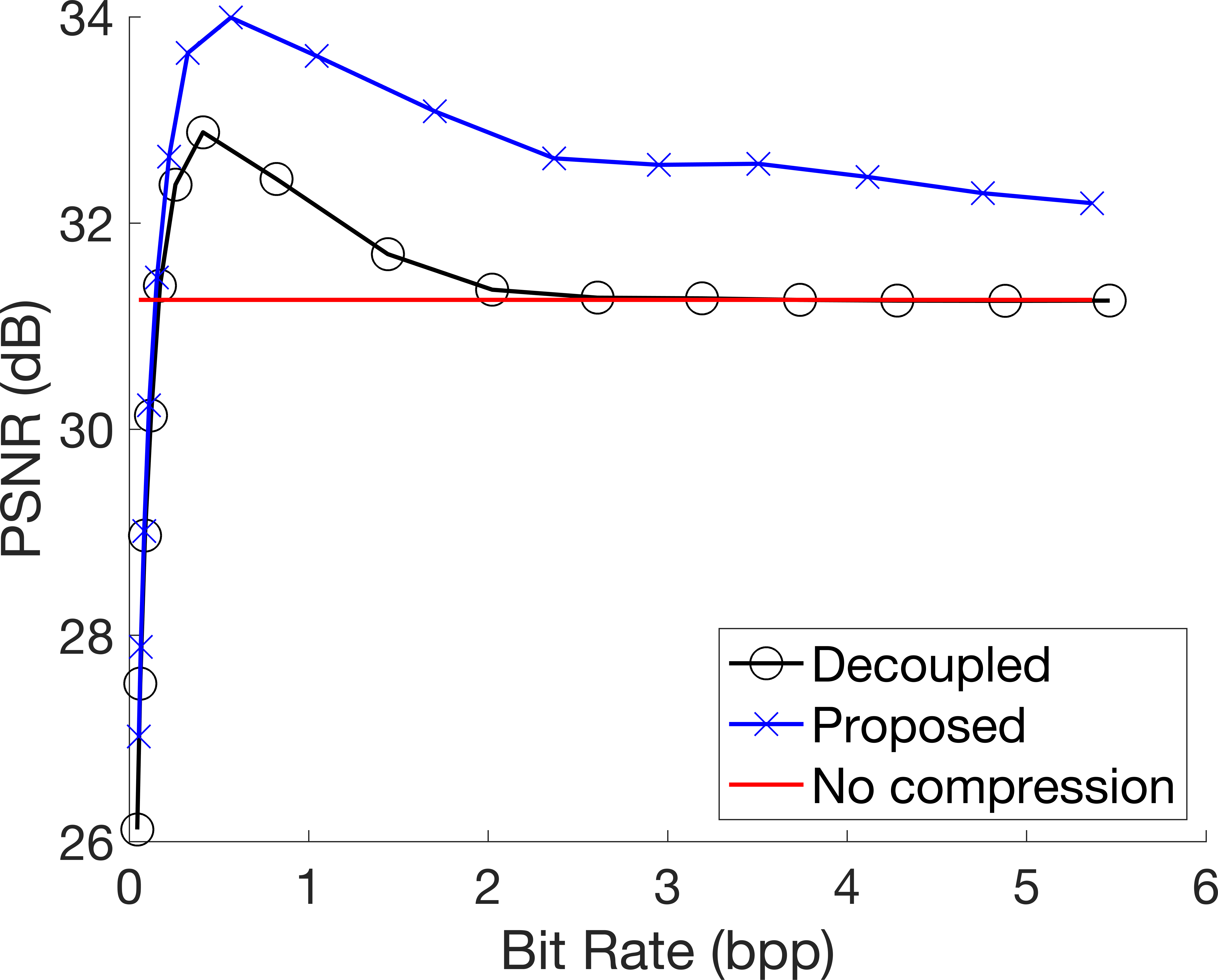}}~
\subfloat[(d) 8x]{\includegraphics[width=0.22\textwidth,]{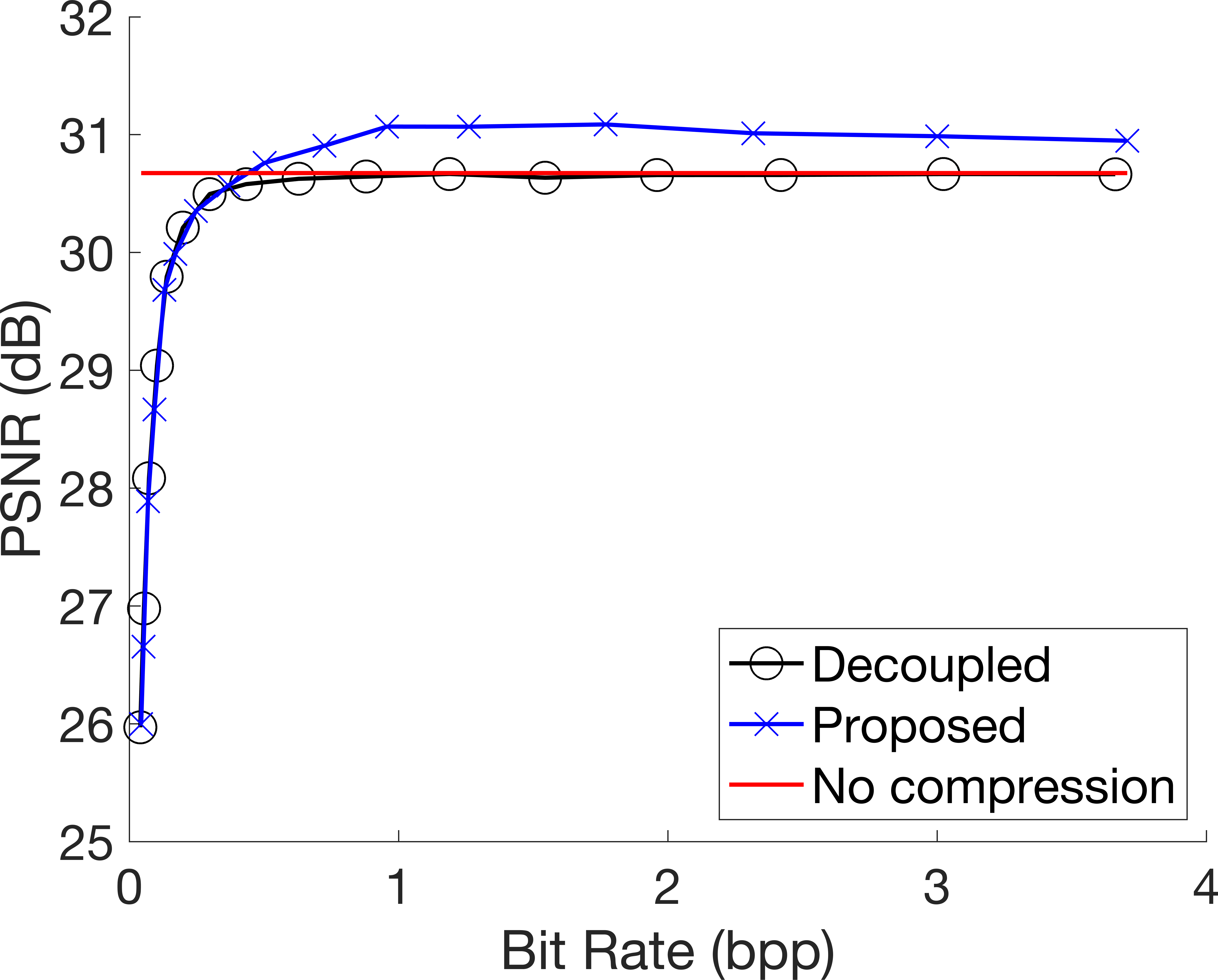}}
\caption{PSNR-bitrate curves comparing the proposed joint optimization using TV-regularization (blue lines) to the decoupled MRI reconstruction and compression using TV-regularization (black lines), and to no compression, only TV-regularized MRI reconstruction (red lines). All results are for the dataset Liver and regularization parameter $\alpha=0.01$. Each subfigure is for a different acceleration factor.}
\label{fig:option2_liver}
\end{figure*}

\begin{figure*}[t!]
    \centering
        \subfloat[]{
   \raisebox{1.5cm}{\rotatebox[origin=t]{90}{\colorbox{gray!10}{With TV Reg.}}} }
\subfloat[(a) Full]{\includegraphics[width=0.22\textwidth]{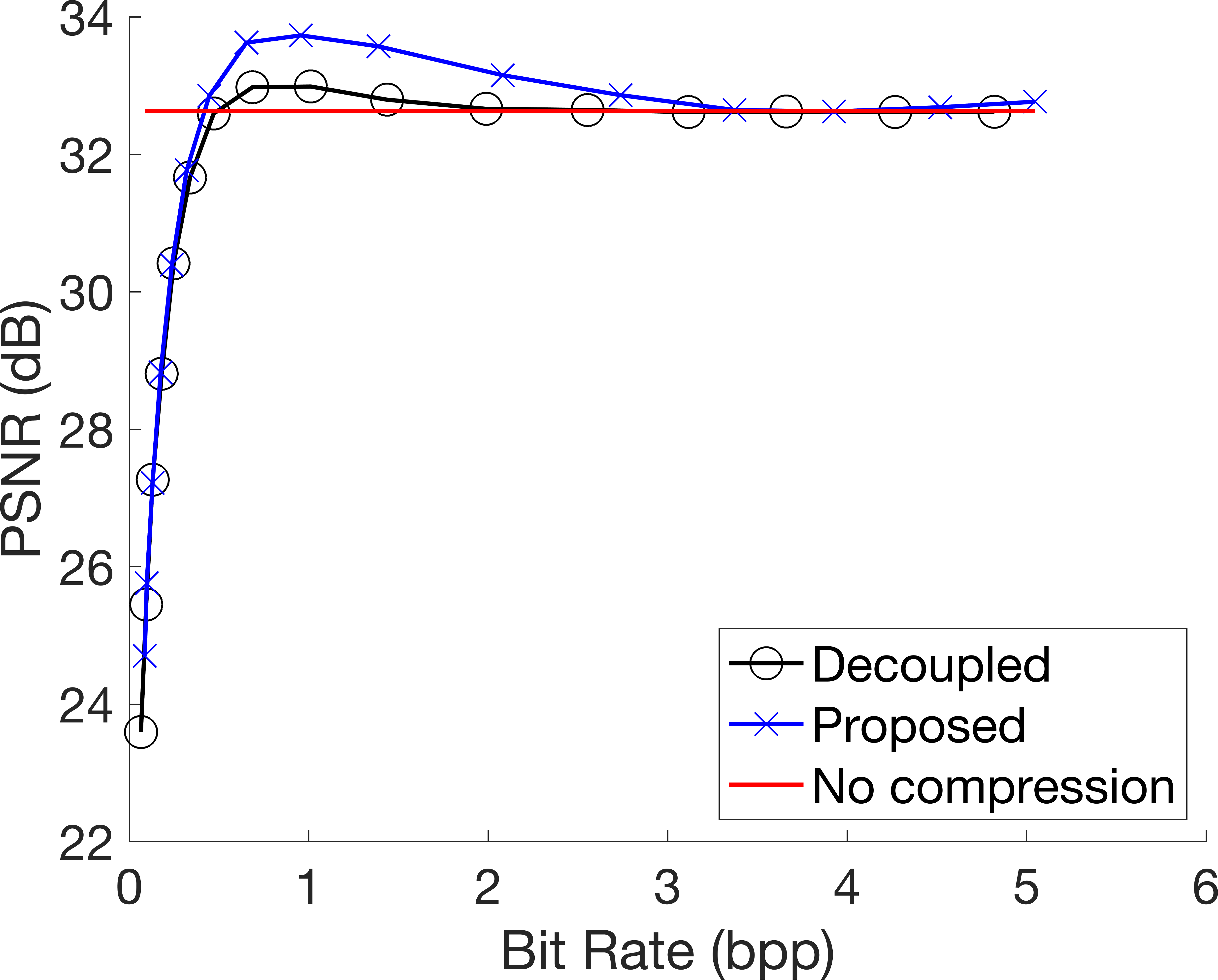}}~
\subfloat[(b) 2x]{\includegraphics[width=0.22\textwidth]{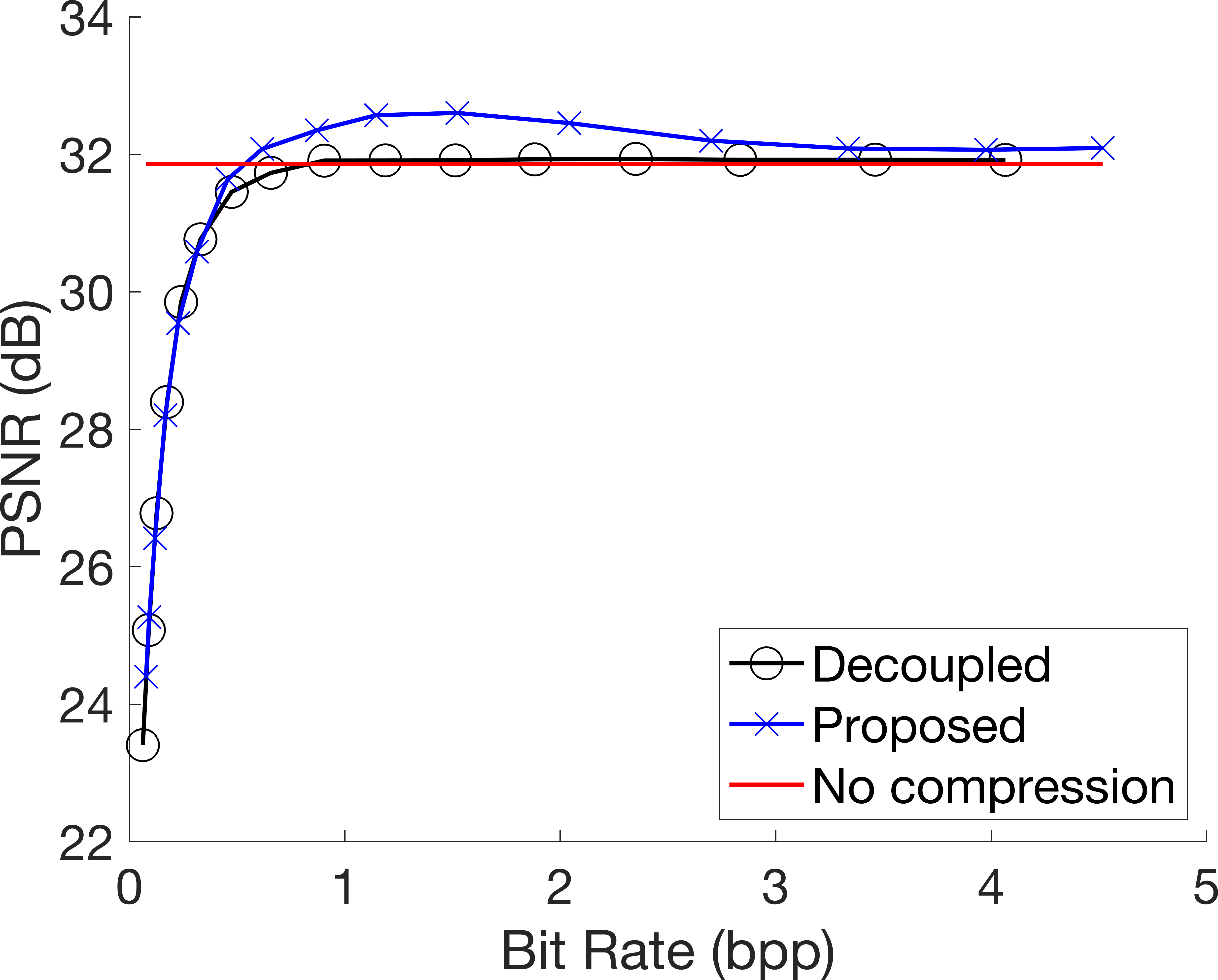}}~
\subfloat[(c) 4x ]{\includegraphics[width=0.22\textwidth,]{figures_psnr/option2_mprage140_25samp_psnr.png}}~
\subfloat[(d) 8x]{\includegraphics[width=0.22\textwidth,]{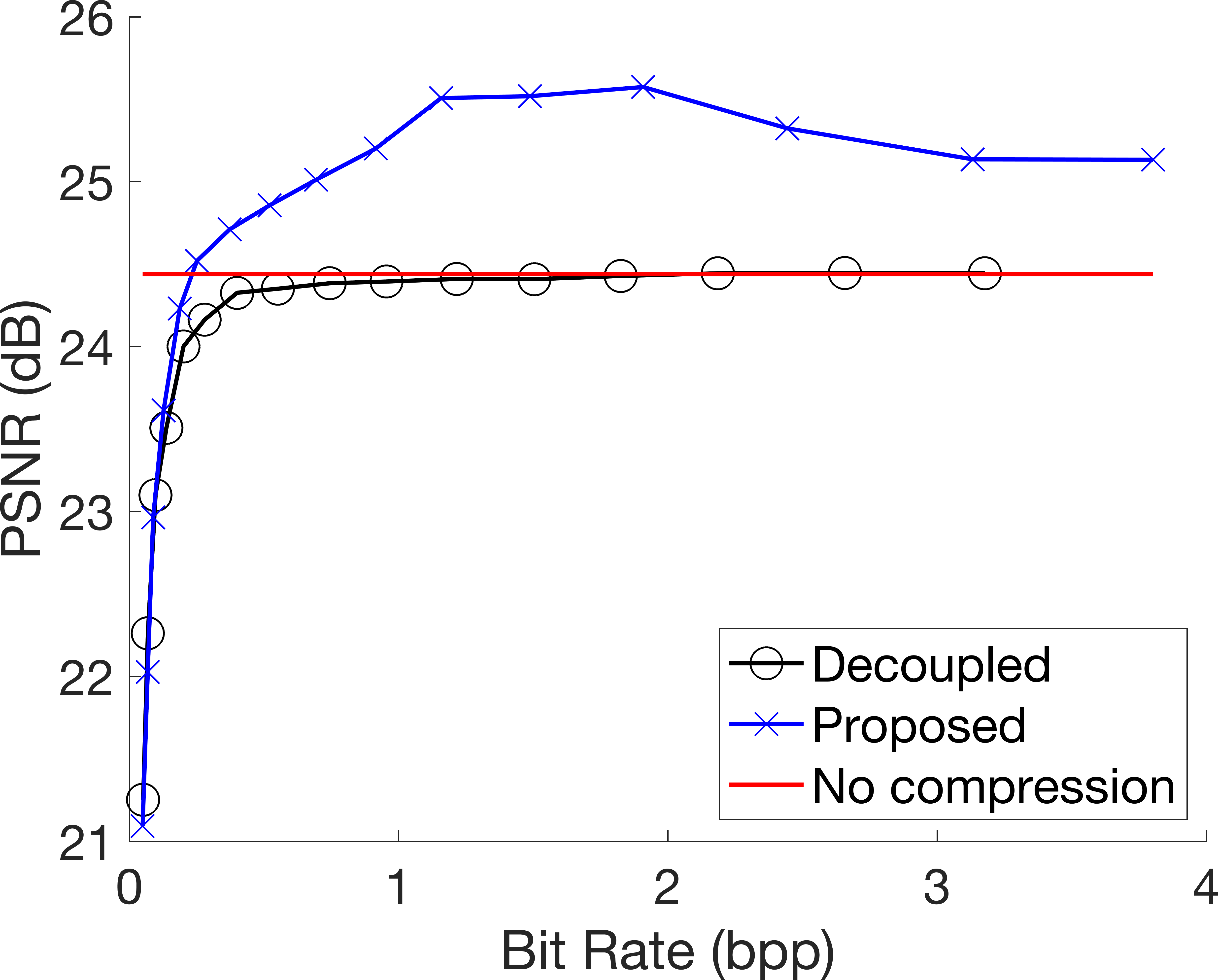}}
\caption{PSNR-bitrate curves comparing the proposed joint optimization using TV-regularization (blue lines) to the decoupled MRI reconstruction and compression using TV-regularization (black lines), and to no compression, only TV-regularized MRI reconstruction (red lines). All results are for the dataset Brain 1 and regularization parameter $\alpha=0.01$. Each subfigure is for a different acceleration factor.}
\label{fig:option2_brain1}
\end{figure*}

Now we proceed to the main part of the results, and evaluate methods with total-variation regularization: 
\begin{itemize}
    \item \textit{Proposed joint optimization with TV-regularization}: This approach is obtained by Algorithm \ref{Algorithm:Proposed Method} with $\alpha>0$. Then, the joint optimization of the MRI reconstruction and lossy compression includes TV-regularization in addition to the implicit regularization provided by the lossy compression. 
    The PSNR versus bit-rate results for this setting are represented by the blue curves in Figures \ref{fig:all_4x_option2}-\ref{fig:option2_brain1}. 

    \item \textit{Decoupled MRI reconstruction and compression with TV-regularization}: in this competing approach, the MR image is reconstructed (using TV regularization) from the measurements, and then compressed by a standard technique (see the general description of the process in Fig.~\ref{fig:decoupled_process}). The PSNR versus bit-rate results for this setting are represented by the black curves in Figures \ref{fig:all_4x_option2}-\ref{fig:option2_brain1}. 
    
    \item \textit{No compression, only MRI reconstruction using TV regularization}: 
    in this setting we measure the PSNR of the MR image reconstructed using TV regularization in a 64-bit numerical resolution. In our comparisons we will refer only to the PSNR values of this setting that are represented by constant red-lines in Figures \ref{fig:all_4x_option2}-\ref{fig:option2_brain1}.
\end{itemize}

The three options that use total-variation regularization provide the PSNR-bitrate curves in Figures \ref{fig:all_4x_option2}-\ref{fig:option2_brain1} for various datasets and acceleration factors. 
These results show that the proposed joint optimization with TV regularization (i.e., Algorithm \ref{Algorithm:Proposed Method} with $\alpha>0$) significantly outperforms, in particular in the medium and high bit-rate range, the two other alternatives (that by themselves utilize TV regularization, as explained above). 

The impressive PSNR gains are summarized in Table \ref{table} that shows the average PSNR difference between performance curves. The average PSNR differences between curves were calculated using the BD-PSNR metric \cite{bjontegaard2001calculation,BDPSNR_Matlab}, for the entire bit-rate range (i.e., the complete curves generated for QP values) and for curve segments corresponding to high bit-rates (defined by QP values 3, 7, 13 and 19). 
Specifically, the results in Table \ref{table} show that our joint optimization method (with TV regularization) is able to achieve PSNR gains of 1 dB at high bit-rates over the decoupled approach that utilizes TV regularization. 
Moreover, Table \ref{table} exhibits the importance of the TV regularization to the proposed method: the PSNR gains of Algorithm \ref{Algorithm:Proposed Method} using TV regularization over Algorithm \ref{Algorithm:Proposed Method} without TV regularization (i.e., with $\alpha=0$) range between 4 to 9 dB. This is a strong evidence for the effectiveness of the proposed joint optimization that includes a novel combination of TV regularization and lossy compression. 

\begin{table*}[t!]
	\caption{~~~~~~~~~~~~~ Average PSNR Gains (DB, measured using BD-PSNR) of Algorithm \ref{Algorithm:Proposed Method} with TV-regularization over Two Alternative Methods: Decoupled Reconstruction and Compression with TV-regularization, and Algorithm \ref{Algorithm:Proposed Method} without TV-regularization.}
	\renewcommand{\arraystretch}{1.1}
	\label{table}
	\centering
	\begin{tabular}{|c|c||c|c||c|c|}
		\hline
		\multirow{2}{*}{\bfseries \shortstack{Image}}  & \multirow{2}{*}{\bfseries \shortstack{MRI\\Acceleration\\ Factor }}  & \multicolumn{2}{|c|}{\bfseries \shortstack{All Bit-Rates}} &  \multicolumn{2}{|c|}{\bfseries\shortstack{High Bit-Rates}} \\
		\cline{3-6}
		& & \shortstack{~\\Proposed with TV-Reg.\\over\\Decoupled with TV Reg.} & \shortstack{~\\Proposed with TV-Reg.\\over\\Proposed without TV Reg.} & \shortstack{~\\Proposed with TV-Reg.\\over\\Decoupled with TV Reg.} & \shortstack{~\\Proposed with TV-Reg.\\over\\Proposed without TV Reg.} \\
		\hline\hline				                                       
		Brain 1 & 2x  & 0.2127 & 5.3815 & 0.4311 & 9.3681 \\
		\cline{2-6}	
		& 4x  & 0.3862 & 4.9906 & 0.9572 & 7.1224 \\
		\cline{2-6}	
		& 8x  & 0.5048 & 3.7934 & 0.9414 & 4.5415 \\
		\hline\hline
		Brain 2 & 2x  & 0.2884 & 5.1590 & 0.6876 & 9.1983 \\
		\cline{2-6}	
		& 4x  & 0.2754 & 4.4242 & 1.1127 & 6.6918 \\
		\cline{2-6}	
		& 8x  & 0.4560 & 3.3232 & 0.8553 & 4.0410 \\
		\hline\hline				                                       
			Brain 3 & 2x  & 0.4075 & 4.8432 & 0.999 & 8.8626 \\
		\cline{2-6}	
		& 4x  & 0.4296 & 4.5707 & 1.1753 & 6.6334 \\
		\cline{2-6}	
		& 8x  & 0.4177 & 3.2280 & 1.0315 & 4.1089 \\
		\hline\hline		     
			Liver & 2x  & -0.1110 & 5.0050 & 0.1190 & 9.8255 \\
		\cline{2-6}	
		& 4x  & 0.1247 & 4.3601 & 0.4618 & 7.0811 \\
		\cline{2-6}	
		& 8x  & 0.0262 & 3.1015 & 0.1728 & 4.0402 \\
		\hline\hline		     
		\hline		                                               
	\end{tabular}
\end{table*}


\begin{figure*}
    \centering
\subfloat[(a) Ground truth]{\includegraphics[width=0.22\textwidth]{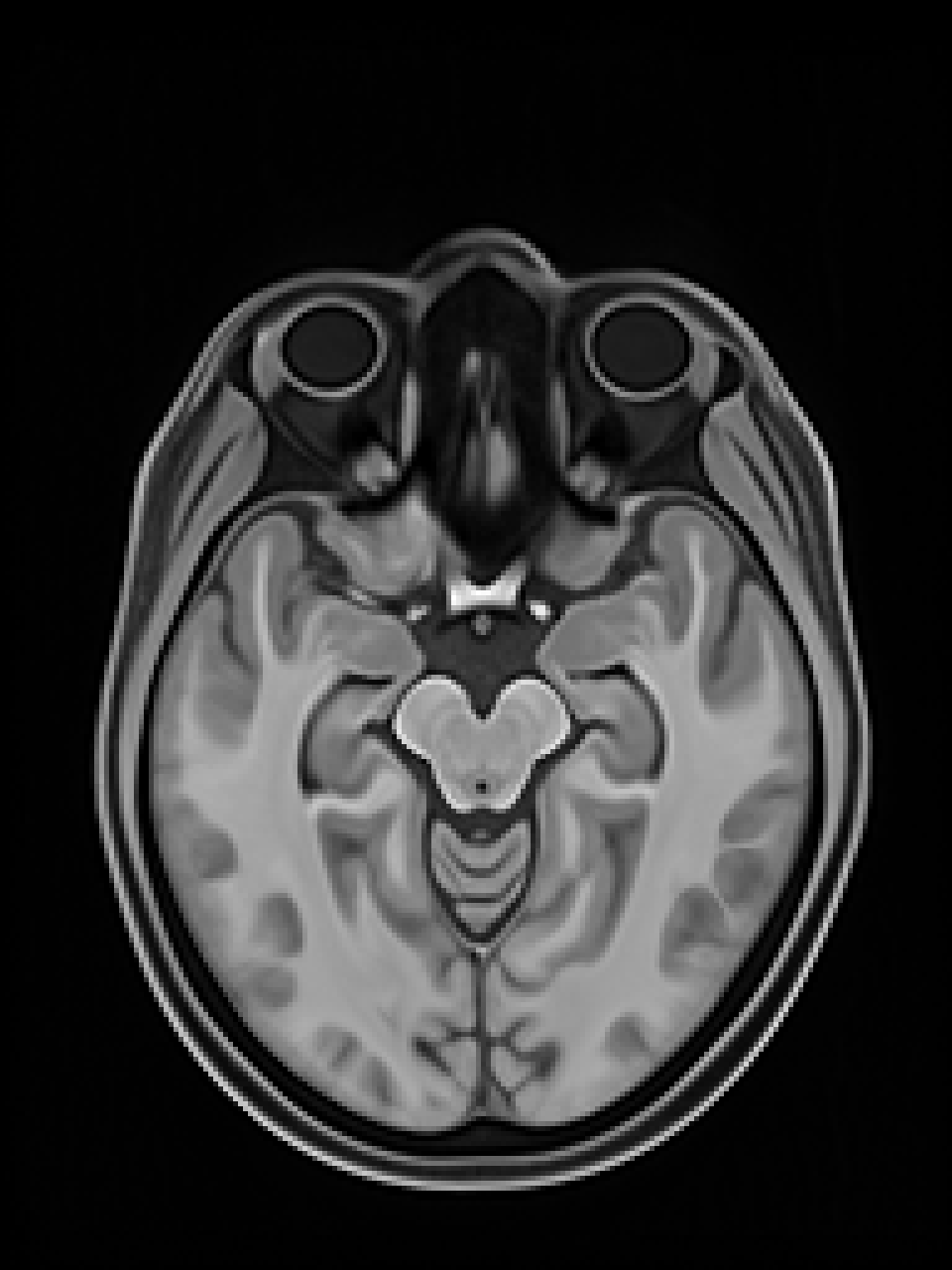}}~
\subfloat[(b) No compression, only TV-reg.~reconstruction \protect\\ PSNR=32.56]{\includegraphics[width=0.22\textwidth]{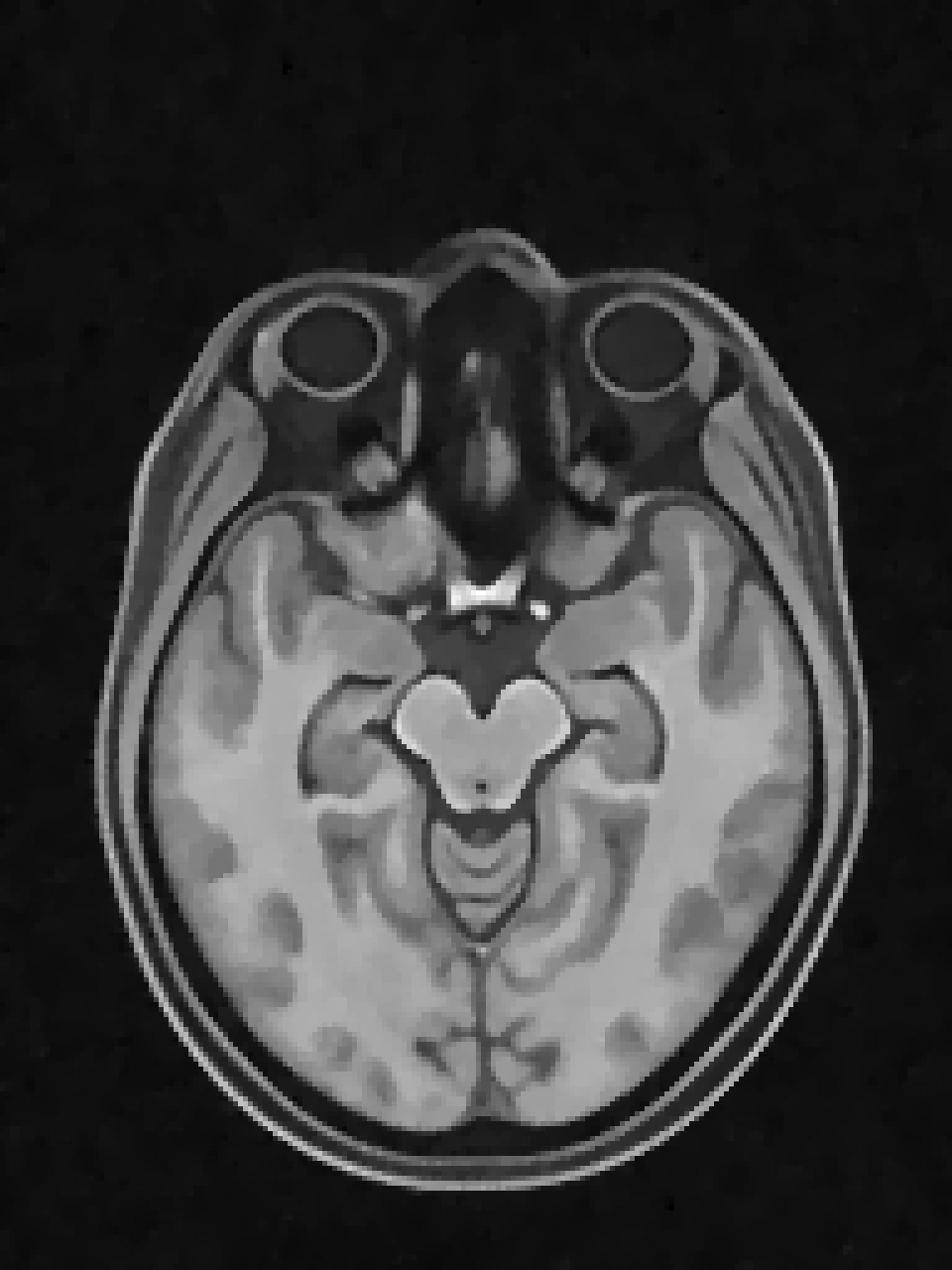}}~
\subfloat[(c) Decoupled with TV reg. \protect\\ PSNR=32.50]{\includegraphics[width=0.22\textwidth,]{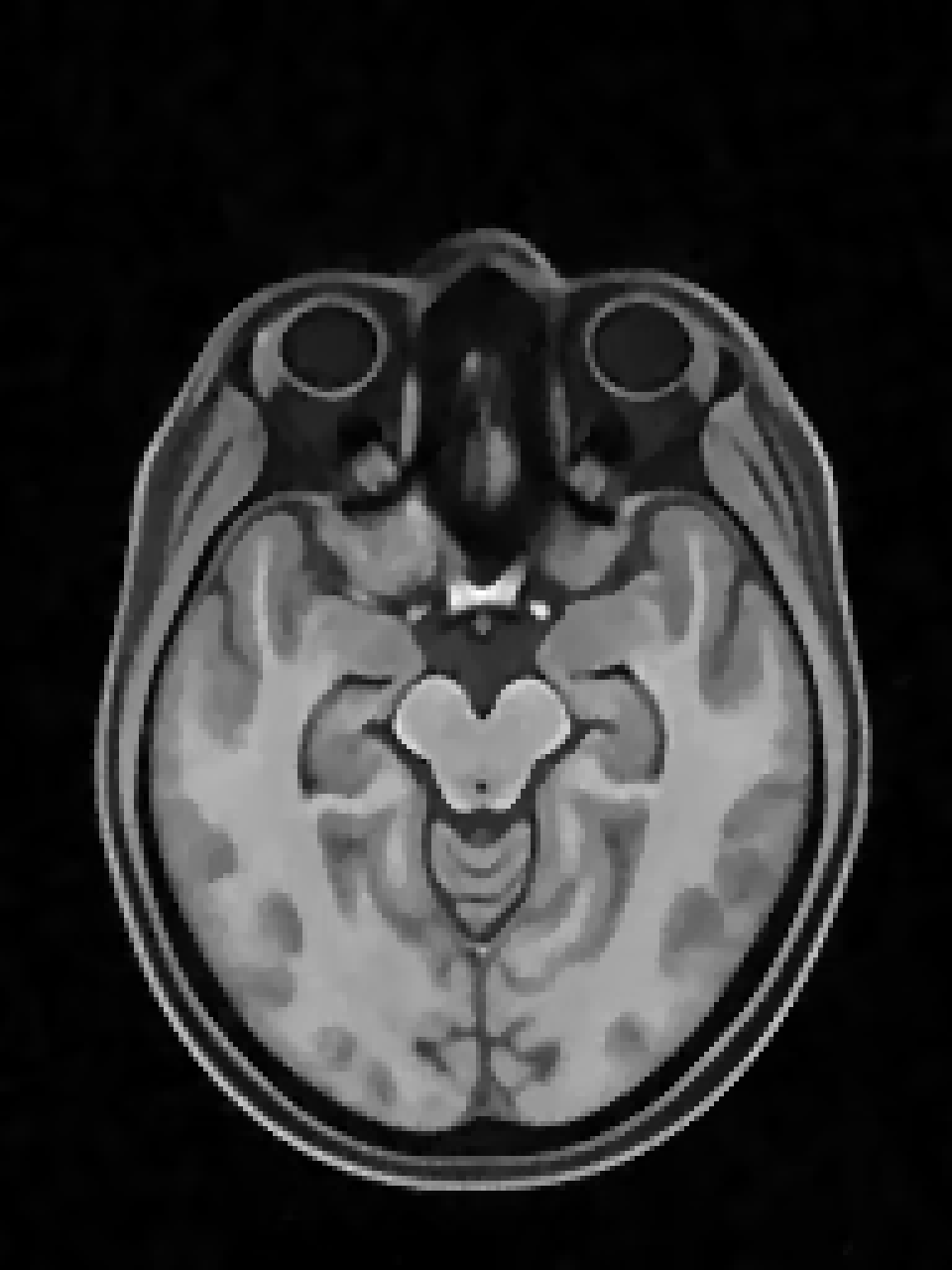}}~
\subfloat[(d) Proposed with TV reg. \protect\\ PSNR=33.74]{\includegraphics[width=0.22\textwidth,]{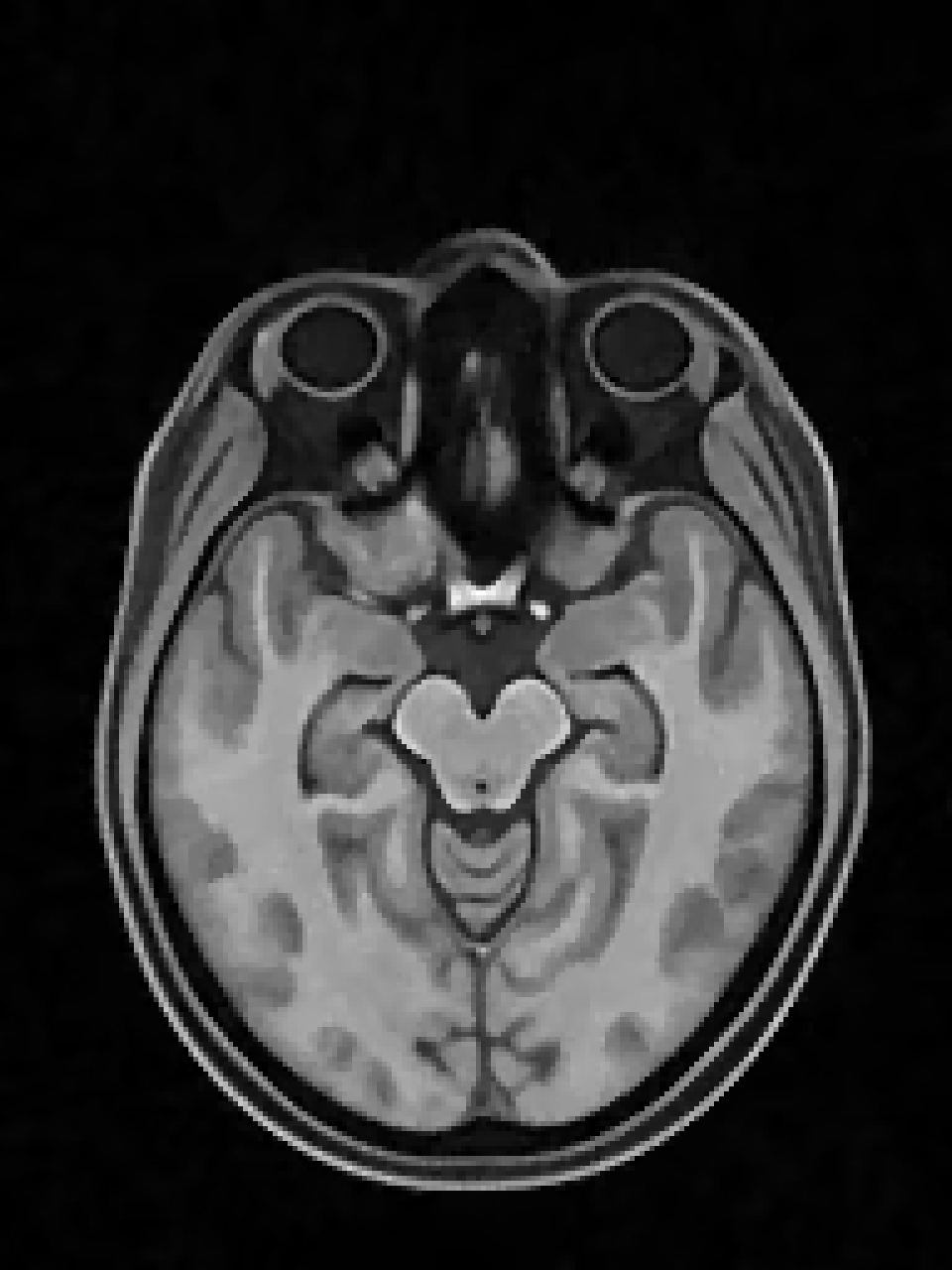}}
\caption{Visual results for data Brain 3 with compression QP 16 for regularized optimization and acceleration factor 2x. We show the (a) ground truth image, (b) no-compression TV-regularized reconstruction ($\alpha=0.02$), (c) decoupled MRI reconstruction and compression \textit{using} TV-regularization ($\alpha=0.02$), and (d) the proposed joint optimization \textit{using} TV-regularization ($\alpha=0.01$). }
\label{fig:visual_option2_brain3}
\end{figure*}

In \autoref{fig:visual_option2_brain3} we provide visual results for the Brain 3 image for the methods that utilize TV regularization. These results clearly show the gains in visual quality (and PSNR) obtained by adding the TV regularization to the lossy compression.

\section{Conclusion}
\label{sec:Conclusion}
In this paper we proposed a new modular optimization method for joint reconstruction and lossy compression of MRI data. The method addresses the degradations in the MRI acquisition stage and provides a compressed representation compatible with a compression standard of choice. We demonstrate that lossy compression can improve the reconstruction quality compared to settings based on lossless compression. An additional novelty is the consideration of a total variation regularizer at the compression stage, leading to a decompressed image of a better quality without any processing at (or after) the decompression stage. 
The proposed method significantly outperforms the relevant competing methods at medium and high bit-rates, showing strong potential for clinical applications. 
Future research directions may explore our approach in conjunction with other regularizers, as well as with automated parameter optimization (e.g., using bi-level optimization \cite{calatroni2017bilevel}). 


\ifCLASSOPTIONcaptionsoff
  \newpage
\fi



\bibliographystyle{IEEEtran}
\bibliography{IEEEabrv,main__refs}
%

%
%

%





\end{document}